\documentclass[12pt]{article}

\usepackage[a4paper,text={16.8cm,22.4cm}]{geometry}
\usepackage{amsmath,amsfonts,slashed,amssymb,tikz,bm,psfrag,graphicx,color,dsfont}
\usepackage{multicol}
\usepackage{float}
\usepackage{multirow}
\usepackage{xcolor}

\RequirePackage[sort&compress,square,comma,numbers]{natbib}
\allowdisplaybreaks
\newcommand{\nn}{\nonumber}

\DeclareMathOperator{\Tr}{Tr}

\begin{document}

\begin{titlepage}

\begin{flushright}
\normalsize
\today
\end{flushright}

\begin{center}
\Large\bf
Higher-twist corrections to $D\to\pi,K$ form factors from light-cone sum rules
\end{center}

\vspace{0.1cm}
\vspace{0.5cm}
\begin{center}
{\bf Long-Sheng Lu$^\ast$,\,Yong-Kang Huang$^\dag$,\,Xuan-Heng Zhang$^\ddag$} \\
\vspace{0.7cm}
{ School of Physics, Nankai University, 300071 Tianjin, China}\\
{Email:\, $^\ast$lulongsheng@mail.nankai.edu.cn,\,$^\dag$2120190131@mail.nankai.edu.cn, \,\\ $^\ddag$1810320@mail.nankai.edu.cn.}
\end{center}

\vspace{0.2cm}

\begin{abstract}
We calculate the $D\to P$ transition form factors within the framework of the light-cone QCD sum rules (LCSR)  with the $D$-meson light-cone distribution amplitudes (LCDAs).   The next-to-leading power (NLP) corrections to the  vacuum-to-$D$-meson correlation function are considered, and the NLP corrections from the high-twist $D$-meson LCDAs and the SU(3) breaking effect from strange quark mass are investigated.
Adopting the exponential model of the $D$-meson LCDAs,
the predicted SU(3) flavor symmetry breaking effects are $R_{SU(3)}^{+,0}=1.12$ and $R_{SU(3)}^{T}=1.39$, respectively, which confirms the results from LCSR with pion LCDA. The numerical predictions of the form factors show that the contribution from two-particle higher-twist contributions is of great importance and the uncertainties are dominated by the inverse moment of $\phi_D^+(\omega, \mu)$.
With the obtained form factors, the predicted Cabibbo-Kobayashi-Maskawa (CKM) matrix elements are $|V_{cd}|=0.151\,{}^{+0.091}_{-0.043} \big |_{\rm th.}\,{}^{+0.017}_{-0.02} \big |_{\rm exp.}$ and $|V_{cs}|=0.89\,{}^{+0.467}_{-0.234} \big |_{\rm th.}\,{}^{+0.008}_{-0.008} \big |_{\rm exp.}$.
\end{abstract}

\vfil

\end{titlepage}

\newpage
\section{Introduction}
The CKM matrix elements are important  parameters of the standard model, which could be extracted from the pure- and semi-leptonic heavy meson decay processes. The semi-leptonic processes have more accurate experimental data since they are not subject to helicity suppression. Recent measurements \cite{Ablikim:2015ixa,Ablikim:2017lks,Ablikim:2018evp} of the semi-leptonic decays $D\to K\,e\,\nu_e$, $D\to K\,\mu\,\nu_{\mu}$, and $D\to \pi\,e\,\nu_e$ provide us with updated branching fractions so that the products of form factors and the CKM matrix $f_{D\to P}(0)|V_{cq}|$ can be extracted directly.  Therefore, the precision prediction of the $D\to P$ form factors is essential in the determination of the CKM matrix elements from the semi-leptonic decay processes.

Theoretically, the heavy-to-light form factors  could be calculated with nonperturbative methods, such as the lattice QCD \cite{Wilson:1974sk},  QCD sum rules \cite{Shifman:1978by}, light-cone sum rules \cite{Balitsky:1989ry,Braun:1988qv,Chernyak:1990ag} methods and quark models \cite{Soni:2017eug,Ivanov:2019nqd,Faustov:2019mqr,Dai:2018vzz}, etc..
The lattice QCD approach is a  QCD-based non-perturbative numerical method which is more suitable to study the processes with small energy release. Thus, phenomenological models are required to extrapolate the  lattice results to the whole kinematic region. Lattice simulation has been applied to $D\to P$ transition form factors in \cite{Na:2010uf,Na:2011mc,Lubicz:2017syv,Lubicz:2018rfs,Li:2019phv}. The QCD sum rules approach is based on local operator product expansion (OPE) and quark hadron duality ansatz, and has been widely used in the hadron physics. To evaluate the heavy-to-light form factors, one must employ the 3-point QCD sum rules \cite{Ball:1998tj,Becirevic:1999kt,Grozin:1996pq,Neubert:1991sp,Ball:1991bs}. In the large recoil region, the light-cone sum rules (LCSR) approach is more appropriate. The nonperturbative input in the LCSR is the LCDAs of hadrons, and both the LCDAs of the initial state hadron and the final state hadron can be employed. The $D\to P$ form factors have been studied with light meson LCSR in \cite{Khodjamirian:2000ds,Ball:2006yd,Khodjamirian:2009ys}.
The latest predictions on the CKM matrix elements $|V_{cd}|=0.225\pm0.005\pm0.003_{-0.012}^{+0.016}$ and $|V_{cd}|/|V_{cs}|=0.236\pm0.006\pm0.003\pm0.013$ \cite{Khodjamirian:2009ys} are in agreement with the CLEO data $|V_{cd}|=0.234\pm0.007\pm0.002\pm0.025$ and $|V_{cs}|=0.985\pm0.009\pm0.003\pm0.103$ \cite{CLEO:2009svp}, which indicates the leading order predictions are reliable.
In this work, we will apply the $D$-meson LCSR to evaluate  the $D\to P$ form factors, and the advantage of our method  is that LCDAs used  in the calculation are universal for different final states.

The idea of $D$-meson LCSR is borrowed from the $B$-meson LCSR \cite{Khodjamirian:2005ea,DeFazio:2005dx}, which has been widely used in heavy flavour physics to investigate the $B\to V$, $B\to P$, $\Lambda_b$ decay processes, etc.,  such as \cite{Gao:2019lta,Wang:2017jow,Shen:2016hyv,Wang:2015ndk}, where QCD corrections and power suppressed contributions have been investigated.  When this approach is applied to the $D$-meson decays, one may worry about the reliability of the light-cone OPE since the energy release is much smaller, actually the light-cone OPE is totally feasible since the the momentum of the interpolation current of the final state is space-like. The $D$-meson LCDA is the nonperturbative input in the $D$-meson LCSR, therefore the predicted form factors can be employed to determine the D-meson LCDA.  The heavy meson LCDA is defined under the framework of the heavy quark effective theory (HQET), where the heavy quark expansion is used. Since the charm quark mass is not large enough, thus the leading power predictions is less accurate. To overcome this shortcomings, we include the power suppressed contributions. In the present paper, we consider the contribution from the higher-twist $D$-meson LCDAs, including the two-particle and three-particle cases.
Although the precision of our predictions is limited, the reliability of our predictions will be improved due to the inclusion of power suppressed contributions. Of course the power suppressed contributions considered in this work is not complete, they can at least provide some instructive results.  With the predicted form factors at hand, we can constrain the CKM matrix elements $|V_ {cq}|$.

This paper is structured as follows. In Sect. 2, we present the formalism of the QCD sum rules and provide the leading-twist contributions to the LCSRs for $D \to \pi, K$ form factors at $\mathcal{O}(\alpha_s)$. In Sect. 3, the higher-twist contributions from two- and three-particle $D$-meson LCDAs are studied. In Sect. 4, we perform the numerical analysis with the predicted form factors from two different $D$-meson LCDA models. Sect. 5 will be reserved for our conclusions.

\section{The $D$-meson LCSR at leading power}
We start with the vacuum-to-$D$-meson correlation function defined with the matrix element of the time-ordered product of a pseudoscalar meson interpolating current and a weak current in the following
 \begin{eqnarray}
\Pi_{\mu}(n \cdot p,\bar n \cdot p)&=& \int d^4x ~e^{i p\cdot x}
\langle 0 |T\left\{\bar{d}(x) \not n \, \gamma_5 \, q(x), \,\,
\bar{q}(0) \, \Gamma_\mu  \, c(0) \right\}|\bar D(p+q) \rangle \nonumber \\
&=&   \left\{
\begin{array}{l}
\Pi(n \cdot p,\bar n \cdot p) \,  n_\mu +\widetilde{\Pi}(n \cdot p,\bar n \cdot p) \, \bar n_\mu \,, \qquad
\Gamma_{\mu}  =  \gamma_{\mu} \vspace{0.4 cm} \\
\Pi_T(n \cdot p,\bar n \cdot p) \, \left [n_{\mu} - \frac{n \cdot q}  {m_D} \, \bar n_{\mu} \right ] \,.
 \qquad  \hspace{1.0 cm} \Gamma_{\mu}  =  \sigma_{\mu \nu} \, q^{\mu}
\end{array}
 \hspace{0.5 cm} \right.
\,
\label{correlator: definition}
\end{eqnarray}
In the above equation, $p+q\equiv m_Dv$ is the four-momentum of the $D$-meson, and $p$ is the momentum of the light pseudoscalar meson.  We  introduce two light-cone vector $\bar{n}_{\mu}$ and $n_{\mu}$ that satisfy $n^2=\bar{n}^2=0$ and $n\cdot \bar{n}=2$, and employ the power counting scheme of light pseudoscalar meson four-momentum which works at the heavy quark limit
\begin{equation}\label{powercounting}
  n\cdot p\sim{\cal O}(m_D),\,\,\,\bar{n}\cdot p\sim m_s\sim{\cal O}(\Lambda),
\end{equation}
where $\Lambda$ above refers to the QCD energy scale. To guarantee the QCD factorization of the correlation function, we will work in the heavy quark limit in the calculation of the leading power contribution. Then the charm quark mass is regarded as a hard scale, and $p$ is hard-collinear, which is similar to $B\to P,V$ transitions. The leading power factorization formulae for the correlation function in $\Lambda/m_c$ are given by
\begin{eqnarray}
\Pi &=& \tilde{f}_D(\mu) \, m_D \sum \limits_{k=\pm} \,
C^{(k)}(n \cdot p, \mu) \, \int_0^{\infty} {d \omega \over \omega- \bar n \cdot p}~
J^{(k)}\left({\mu^2 \over n \cdot p \, \omega},{\omega \over \bar n \cdot p}\right) \,
\phi_D^{k}(\omega,\mu)  \,,
\nonumber \\
\widetilde{\Pi} &=& \tilde{f}_D(\mu) \, m_D \sum \limits_{k=\pm} \,
\widetilde{C}^{(k)}(n \cdot p, \mu) \, \int_0^{\infty} {d \omega \over \omega- \bar n \cdot p}~
\widetilde{J}^{(k)}\left({\mu^2 \over n \cdot p \, \omega},{\omega \over \bar n \cdot p}\right) \,
\phi_D^{k}(\omega,\mu)  \,,
\nonumber \\
\Pi_T &=& - {i \over 2} \,  \tilde{f}_D(\mu) \, m_D^2 \,  \sum \limits_{k=\pm} \,
C^{(k)}_T(n \cdot p, \mu, \nu) \, \int_0^{\infty} {d \omega \over \omega- \bar n \cdot p}~
J^{(k)}_T \left({\mu^2 \over n \cdot p \, \omega},{\omega \over \bar n \cdot p}\right) \,
\phi_D^{k}(\omega,\mu) \,,
\label{QCD factorization formula of correlator at LT}
\end{eqnarray}
where functions $C^{(k)}$ and $J^{(k)}$ are hard and jet functions, respectively. The $D$-meson LCDAs are defined by the following renormalized matrix element \cite{Grozin:1996pq}
\begin{eqnarray}
&& \langle  0 | \left ( \bar{u}\, Y_s^{\dag} \right )_{\alpha}(\tau \, \bar{n}) \,\left ( Y_s^{\dag}\,h_v  \right)_{\beta} (0)
| D(v)\rangle \nonumber \\
&& = - \frac{i \tilde f_D(\mu) \, m_D}{4}  \bigg \{ \frac{1+ \! \not v}{2} \,
\left [ 2 \, \tilde{\phi}_{D}^{+}(\tau, \mu) + \left ( \tilde{\phi}_{D}^{-}(\tau, \mu)
-\tilde{\phi}_{D}^{+}(\tau, \mu)  \right )  \! \not n \right ] \, \gamma_5 \bigg \}_{\alpha \beta}\,.
\label{def: two-particle D-meson DAs}
\end{eqnarray}
In the above equation, $Y_s(\tau \bar{n})$ is the soft Wilson line ensuring the gauge invariance
\begin{eqnarray}
Y_s(\tau \, \bar n)= {\rm P} \, \left \{ {\rm  Exp} \left [   i \, g_s \,
\int_{- \infty}^{\tau} \, dx \,  \bar n  \cdot A_{s}(x \, \bar n) \right ]  \right \} \,,
\label{def: soft gauge link}
\end{eqnarray}
and $\tilde f_D(\mu)$ is the renormalization-scale dependent HQET decay constant
\begin{eqnarray}
\tilde{f}_D(\mu)= \left \{  1 -  {\alpha_s(\mu) \, C_F \over 4 \, \pi} \,
\left [3\, \ln{\mu \over m_c} + 2  \right ] \right \}^{-1} \, f_D \,.
\label{HQET matching of fD}
\end{eqnarray}
At one-loop accuracy, the hard and jet functions are given by \cite{Lu:2018cfc}
\begin{eqnarray}
C^{(+)} &=&  \widetilde{C}^{(+)} = C_T^{(+)} = 1 \,, \qquad
C^{(-)} =  {\alpha_s \, C_F \over 4 \pi} \, {1 \over \bar r} \,
\left [ 1 + { r \over \bar r} \, \ln r \right ] \,, \nonumber \\
\widetilde{C}^{(-)} &=&  1-  {\alpha_s \, C_F \over 4 \pi} \,
\left [ 2 \, \ln^2 {\mu \over n \cdot p} + 5 \, \ln {\mu \over m_b}
- \ln^2 r - 2 \, {\rm Li}_2 \left (- { \bar r \over r } \right )
+ { 2 - r \over r-1}  \, \ln r + {\pi^2 \over 12} + 5 \right ] \,, \nonumber \\
C_T^{(-)} &=&   1 +  {\alpha_s \, C_F \over 4 \pi} \,
\left [ -2 \, \ln {\nu \over m_b} - 2 \, \ln^2 {\mu \over n \cdot p}
- 5 \, \ln {\mu \over n \cdot p}  - 2 \, {\rm Li}_2 (1-r)
- {3 - r \over 1 -r} \, \ln r -  {\pi^2 \over 12} - 6 \right ] \,, \nonumber \\
J^{(+)} &=&   {\alpha_s \, C_F \over 4 \pi} \, \left (1 -  {\bar n \cdot p \over \omega} \right ) \,
\ln \left (1- {\omega \over \bar n \cdot p} \right ) \,, \nonumber \\
\widetilde{J}^{(+)} &=&   {\alpha_s \, C_F \over 4 \pi} \,
\left [ r \, \left (1 -  {\bar n \cdot p \over \omega} \right ) + {m_q \over \omega} \right ] \,
\ln \left (1- {\omega \over \bar n \cdot p} \right ) \,, \nonumber  \\
J_T^{(+)} &=& {\alpha_s \, C_F \over 4 \pi} \,
\left [ - \, \left (1 -  {\bar n \cdot p \over \omega} \right ) + {m_q \over \omega} \right ] \,
\ln \left (1- {\omega \over \bar n \cdot p} \right )  \,, \nonumber  \\
J^{(-)} &=&  1  \,, \nonumber  \\
\widetilde{J}^{(-)} &=&   J_T^{(-)}
=  1 + \frac{\alpha_s \, C_F}{4 \, \pi} \,
\bigg [ \ln^2 { \mu^2 \over  n \cdot p (\omega- \bar n \cdot p) }
- 2 \ln {\bar n \cdot p -\omega \over \bar n \cdot p } \, \ln { \mu^2 \over  n \cdot p (\omega- \bar n \cdot p) }
\,  \nonumber \\
&& - \ln^2 {\bar n \cdot p -\omega \over \bar n \cdot p }
- \left ( 1 +  {2 \bar n \cdot p \over \omega} \right )  \ln {\bar n \cdot p -\omega \over \bar n \cdot p }
-{\pi^2 \over 6} -1 \bigg ] \,,
\label{matching coefficients of the LP contributions}
\end{eqnarray}
where $\nu$ is the renormalization scale of the tensor current. Conventions of $r$ and $\bar{r}$ are $r=n\cdot p/m_c$ and $\bar{r}=1-r$, respectively.

To obtain the next-to-leading-logarithmic (NLL) accuracy factorization formulae, we employ the renormalization group equation in the momentum space and take the factorization scale $\mu$  as a hard-collinear scale $\mu_{hc}\sim\sqrt{\Lambda_{\rm QCD}\,m_c}$. By solving the evolution equations, one obtain the NLL resummation improved expressions for the hard function and the HQET decay constant
\begin{eqnarray}
\widetilde{C}^{(-)}(n \cdot p, \mu) &=& U_1(n \cdot p, \mu_{h1}, \mu) \,\, \widetilde{C}^{(-)}(n \cdot p, \mu_{h1})\,, \nonumber \\
C^{(-)}_{T}(n \cdot p, \mu, \nu)  &=& U_1(n \cdot p, \mu_{h1}, \mu) \,\, U_3(\nu_{h}, \nu) \,\,
C^{(-)}_{T}(n \cdot p, \mu_{h1}, \nu_{h}) \,,  \nonumber \\
\tilde{f}_D(\mu)&=& \, U_2(\mu_{h2}, \mu) \, \tilde{f}_D(\mu_{h2}) \,\,,
\end{eqnarray}
where the expressions of the evolution functions $U_1$, $U_2$ and $U_3$ are collected in the appendix.
Then the factorization formulae for the correlation functions at NLL accuracy read
\begin{eqnarray}
\Pi &=&  \left [ U_2(\mu_{h2}, \mu) \, \tilde{f}_D(\mu_{h2}) \right ]\, m_D  \,
\, \bigg \{  \int_0^{\infty} {d \omega \over \omega- \bar n \cdot p}~
J^{(+)}\left({\mu^2 \over n \cdot p \, \omega},{\omega \over \bar n \cdot p}\right) \,
\phi_D^{+}(\omega,\mu)  \nonumber \\
&&  +  \, C^{(-)}(n \cdot p, \mu)  \,
\int_0^{\infty} {d \omega \over \omega- \bar n \cdot p}~ \,
\phi_D^{-}(\omega,\mu)  \,  \bigg \}\,,
\nonumber  \\
\widetilde{\Pi} &=&  \left [ U_2(\mu_{h2}, \mu) \, \tilde{f}_D(\mu_{h2}) \right ]\, m_D  \,
\, \bigg \{  \int_0^{\infty} {d \omega \over \omega- \bar n \cdot p}~
\widetilde{J}^{(+)}\left({\mu^2 \over n \cdot p \, \omega},{\omega \over \bar n \cdot p}\right) \,
\phi_D^{+}(\omega,\mu)  \nonumber \\
&&  +  \,\left [  U_1(n \cdot p, \mu_{h1}, \mu) \,\, \widetilde{C}^{(-)}(n \cdot p, \mu_{h1}) \right ] \,
\int_0^{\infty} {d \omega \over \omega- \bar n \cdot p}~
\widetilde{J}^{(-)}\left({\mu^2 \over n \cdot p \, \omega},{\omega \over \bar n \cdot p}\right) \,
\phi_D^{-}(\omega,\mu)  \,  \bigg \}\,,
\nonumber  \\
\Pi_T &=& - {i \over 2} \, \left [ U_2(\mu_{h2}, \mu) \, \tilde{f}_D(\mu_{h2}) \right ]\, m_D^2  \,
\, \bigg \{  \int_0^{\infty} {d \omega \over \omega- \bar n \cdot p}~
J_T^{(+)}\left({\mu^2 \over n \cdot p \, \omega},{\omega \over \bar n \cdot p}\right) \,
\phi_D^{+}(\omega,\mu)  \nonumber \\
&&  +  \,\left [ U_1(n \cdot p, \mu_{h1}, \mu) \,\, U_3(\nu_{h}, \nu) \,\,
C^{(-)}_{T}(n \cdot p, \mu_{h1}, \nu_{h})  \right ] \, \nonumber  \\
&& \hspace{0.5 cm} \times \, \int_0^{\infty} {d \omega \over \omega- \bar n \cdot p}~
J_T^{(-)}\left({\mu^2 \over n \cdot p \, \omega},{\omega \over \bar n \cdot p}\right) \,
\phi_D^{-}(\omega,\mu)  \,  \bigg \} \,.
\label{Resummation improved factorization formula of correlator at LT}
\end{eqnarray}

The definitions of $D\to P$ form factors and the decay constant of the pseudoscalar meson are given by
\begin{eqnarray}
\langle P(p)|  \bar q \, \gamma_{\mu} \, c| \bar D (p+q)\rangle
&=& f_{D \to P}^{+}(q^2) \, \left [ 2 p + q -\frac{m_D^2-m_{P}^2}{q^2} q  \right ]_{\mu}
+  f_{D \to P}^{0}(q^2) \, \frac{m_D^2-m_{P}^2}{q^2} q_{\mu} \,, \nonumber \\
\langle P(p)|  \bar q \, \sigma_{\mu \nu} \, q^{\nu}\, c| \bar D (p+q)\rangle
&=& i \, {f_{D \to P}^{T}(q^2) \over m_D + m_P}\, \left [ q^2   \,\, (2 p + q)_{\mu} \,
-(m_D^2-m_{P}^2) \, q_{\mu}  \right ] \,, \nonumber \\
\langle 0 |\bar d \! \not n \, \gamma_5 \, q |  P(p)  \rangle &=&  i \, n \cdot p \, f_{P} \,,
\end{eqnarray}
where $f_{D \to P}^{+}(q^2)$ and $f_{D \to P}^{T}(q^2)$ are the vector and tensor $D\to P$ form factors, respectively. By inserting the above equations into the hadronic-level vacuum-to-$D$-meson correlation function
\begin{eqnarray}
\Pi_{\mu}(n \cdot p,\bar n \cdot p) &=& \frac{\langle 0|\bar{d}\not n \,\gamma_5\,q|P(p)\rangle\langle P(p)|\bar{q}\,\Gamma_{\mu}\,c|\bar{D}(p+q)\rangle}{m_P^2-p^2}+continuum\, states,
\label{hadronic-level D-meson LCDA}
\end{eqnarray}
we can readily obtain the hadronic representations of the correlation function for the vector and tensor Dirac structures
\begin{eqnarray}
\Pi_{\mu, V}(n \cdot p,\bar n \cdot p) &=& \frac{f_{P} \, m_D}{2 \, (m_{P}^2/ n \cdot p - \bar n \cdot p)}
\bigg \{  \bar n_{\mu} \, \left [ \frac{n \cdot p}{m_D} \, f_{D \to P}^{+} (q^2) + f_{D \to P}^{0} (q^2)  \right ]
\nonumber \\
&& \hspace{0.4 cm} +  \,  n_{\mu} \, \frac{m_D}{n \cdot p-m_D}  \, \,
\left [ \frac{n \cdot p}{m_D} \, f_{D \to P}^{+} (q^2) -  f_{D \to P}^{0} (q^2)  \right ] \bigg \} \, \nonumber \\
&& \hspace{0.4 cm} + \int_{\omega_s}^{+\infty}   \, \frac{d \omega^{\prime} }{\omega^{\prime} - \bar n \cdot p - i 0} \,
\left [ \rho_{V, 1}^{h}(\omega^{\prime}, n \cdot p)  \, n_{\mu} \,
+\rho_{V, 2}^{h}(\omega^{\prime}, n \cdot p)  \, \bar{n}_{\mu}  \right ] \,, \nonumber \\
\Pi_{\mu, T}(n \cdot p,\bar n \cdot p) &=& - i \, \frac{f_{P} \, n \cdot p }{2 \, (m_{P}^2/ n \cdot p - \bar n \cdot p)}
\, {m_D^2 \over m_D+m_P} \, \left [ n_{\mu} -  {n \cdot q \over m_D} \, \bar n_{\mu} \right ]  \, f_{D \to P}^{T} (q^2) \nonumber \\
&& \hspace{0.4 cm} + \int_{\omega_s}^{+\infty}   \, \frac{d \omega^{\prime} }{\omega^{\prime} - \bar n \cdot p - i 0} \,
\, \left [ n_{\mu} -  {n \cdot q \over m_D} \, \bar n_{\mu} \right ]  \,
\rho_{T}^{h}(\omega^{\prime}, n \cdot p) \,.
\label{hadronic representations}
\end{eqnarray}
To obtain the NLL LCSR for $D \to P$ form factors at leading power in the heavy quark expansion, one can match the HQET representation and the hadronic representation based on the parton-hadron duality ansatz. Implementing the Borel transformation, one obtains
\begin{eqnarray}
&& f_{P} \,\, {\rm exp} \left [- {m_{P}^2 \over n \cdot p \,\, \omega_M} \right ] \,\,
\left \{ \frac{n \cdot p} {m_D} \, f_{D \to P}^{+, \, \rm 2PNLL}(q^2)
\,, \,\,\,   f_{D \to P}^{0, \, \rm 2PNLL}(q^2)  \right \}  \,  \nonumber \\
&& =   \left [ U_2(\mu_{h2}, \mu) \, \tilde{f}_D(\mu_{h2}) \right ]
\,\, \int_0^{\omega_s} \,\, d \omega^{\prime} \, e^{-\omega^{\prime}/\omega_M} \,  \nonumber \\
&&  \hspace{0.4 cm} \times \bigg \{\widetilde{\phi}_{D, \, \rm {eff}}^{+} (\omega^{\prime}, \mu)
+  \, \left [  U_1(n \cdot p, \mu_{h1}, \mu) \,\, \widetilde{C}^{(-)}(n \cdot p, \mu_{h1}) \right ]\,
\widetilde{ \phi}_{D, \, \rm {eff}}^{-} (\omega^{\prime}, \mu) \nonumber \\
&& \hspace{0.8 cm} \pm \, { n \cdot p - m_D \over m_D} \,
\left [\phi_{D, \, \rm {eff}}^{+} (\omega^{\prime}, \mu)
+ C^{(-)}(n \cdot p, \mu_{h1}) \, \phi_{D, \, \rm {eff}}^{-} (\omega^{\prime}, \mu)   \right ]  \bigg \} \,, \nonumber \\
&& f_{P} \,\,  {\rm exp} \left [- {m_{P}^2 \over n \cdot p \,\, \omega_M} \right ]  \,\,
\frac{n \cdot p} {m_D+m_P} \,  f_{D \to P}^{T, \, \rm 2PNLL}(q^2) \,  \nonumber \\
&& =   \left [ U_2(\mu_{h2}, \mu) \, \tilde{f}_D(\mu_{h2}) \right ]
\,\, \int_0^{\omega_s} \,\, d \omega^{\prime} \, e^{-\omega^{\prime}/\omega_M} \,  \nonumber \\
&&  \hspace{0.4 cm} \times \bigg \{ \widehat{\phi}_{D, \, \rm {eff}}^{+} (\omega^{\prime}, \mu)
+  \,\left [ U_1(n \cdot p, \mu_{h1}, \mu) \,\, U_3(\nu_{h}, \nu) \,\,
C^{(-)}_{T}(n \cdot p, \mu_{h1}, \nu_{h})  \right ] \,
\widetilde{ \phi}_{D, \, \rm {eff}}^{-} (\omega^{\prime}, \mu)  \bigg \} \,.
\label{NLO sum rules of D to P form factors}
\end{eqnarray}
The effective $D$-meson DAs are defined as \cite{Lu:2018cfc}
\begin{eqnarray}
\widetilde{\phi}_{D, \, \rm {eff}}^{+} (\omega^{\prime}, \mu) &=&
{\alpha_s \, C_F \over 4 \, \pi} \,
\left [ r \, \int_{\omega^{\prime}}^{\infty} \, d \omega \,
{\phi_D^{+}(\omega, \mu) \over \omega}
- m_q \, \int_{\omega^{\prime}}^{\infty} \, d \omega \,
\ln \left ( {\omega -  \omega^{\prime} \over \omega^{\prime}} \right )   \,\,
{d \over d \omega} \, {\phi_D^{+}(\omega, \mu) \over \omega}  \right ] \,,  \nonumber \\
\widetilde{\phi}_{D, \, \rm {eff}}^{-} (\omega^{\prime}, \mu) &=&
\phi_{D}^{-}(\omega^{\prime}, \mu)
+  \frac{\alpha_s \, C_F}{4 \, \pi} \,\, \bigg \{ \int_0^{\omega^{\prime}} \,\, d \omega \,\,\,
\left [ {2 \over \omega - \omega^{\prime}}  \,\,\, \left (\ln {\mu^2 \over n \cdot p \, \omega^{\prime}}
- 2 \, \ln {\omega^{\prime} - \omega \over \omega^{\prime}} \right )\right ]_{\oplus} \,
\phi_{D}^{-}(\omega, \mu)  \nonumber \\
&& - \int_{\omega^{\prime}}^{\infty} \,\, d \omega \,\,\,
\bigg [ \ln^2 {\mu^2 \over n \cdot p \, \omega^{\prime}}
- \left ( 2 \, \ln {\mu^2 \over n \cdot p \, \omega^{\prime}}  + 3 \right ) \,\,
\ln {\omega - \omega^{\prime} \over \omega^{\prime}}
+ \, 2 \,\, \ln {\omega \over \omega^{\prime}}   + {\pi^2 \over 6} - 1 \bigg ]
\nonumber \\
&& \hspace{0.5 cm}  \times \, {d \phi_{D}^{-}(\omega, \mu) \over d \omega}  \bigg \}  \,,   \nonumber \\
\phi_{D, \, \rm {eff}}^{+} (\omega^{\prime}, \mu) &=&  {\alpha_s \, C_F \over 4 \, \pi} \,
\int_{\omega^{\prime}}^{\infty} \, d \omega \, {\phi_D^{+}(\omega, \mu) \over \omega} \,,
\qquad
\phi_{D, \, \rm {eff}}^{-} (\omega^{\prime}, \mu) = \phi_D^{-}(\omega^{\prime}, \mu) \,, \nonumber \\
\widehat{\phi}_{D, \, \rm {eff}}^{+} (\omega^{\prime}, \mu) &=&
{\alpha_s \, C_F \over 4 \, \pi} \,
\left [ - \int_{\omega^{\prime}}^{\infty} \, d \omega \,
{\phi_D^{+}(\omega, \mu) \over \omega}
- m_q \, \int_{\omega^{\prime}}^{\infty} \, d \omega \,
\ln \left ( {\omega -  \omega^{\prime} \over \omega^{\prime}} \right )   \,\,
{d \over d \omega} \, {\phi_D^{+}(\omega, \mu) \over \omega}  \right ] \,,
\label{effective D-meson DAs}
\end{eqnarray}
where the strange quark mass effect at $\alpha_{s}$ could be observed in the above equation. The plus function entering (\ref{effective D-meson DAs}) is defined as
\begin{eqnarray}
\int_0^{\infty} \, d \omega \, \left [ f(\omega, \omega^{\prime}) \right ]_{\oplus} \, g(\omega)
= \int_0^{\infty} \, d \omega \,  f(\omega, \omega^{\prime})
\left [ g(\omega) - g(\omega^{\prime}) \right ]   \,.
\end{eqnarray}

\section{The higher-twist contributions}
In this section, we will evaluate the higher-twist contributions from two-particle and three-particle $D$-meson LCDAs.
To compute the three-particle higher-twist contributions of $D\to\pi\,,K$ form factors, we apply the light-cone expansion of the quark propagator in the background gluon field \cite{Balitsky:1987bk}
\begin{eqnarray}
\langle 0 | {\rm T} \, \{\bar q (x), q(0) \} | 0\rangle
 \supset   i \, g_s \, \int_0^{\infty} \,\, {d^4 k \over (2 \pi)^4} \, e^{- i \, k \cdot x} \,
\int_0^1 \, d u \, \left  [ {u \, x_{\mu} \, \gamma_{\nu} \over k^2 - m_q^2}
 - \frac{(\not \! k + m_q) \, \sigma_{\mu \nu}}{2 \, (k^2 - m_q^2)^2}  \right ]
\, G^{\mu \nu}(u \, x) \,, \hspace{0.4 cm}
\end{eqnarray}
where only the one-gluon part is considered. The $D$-meson three-particle LCDAs are defined as \cite{Braun:2017liq,Lu:2018cfc}
\begin{eqnarray}
\lefteqn{\langle 0| \bar q(nz_1) g_s\,G_{\mu\nu}(nz_2)\Gamma h_v(0) |\bar D(v)\rangle =}
\nonumber\\
&=&
\frac14 F_B(\mu) \Tr\biggl\{\gamma_5 \Gamma (1 + \not v)
\biggl[ (v_\mu\gamma_\nu-v_\nu\gamma_\mu)  \big[{\Psi}_A-{\Psi}_V \big]-i\sigma_{\mu\nu}{\Psi}_V
- (n_\mu v_\nu-n_\nu v_\mu){X}_A
\nonumber\\&&{}\hspace*{0.1cm}
 + (n_\mu \gamma_\nu-n_\nu \gamma_\mu)\big[W+{Y}_A\big]
- i\epsilon_{\mu\nu\alpha\beta} n^\alpha v^\beta \gamma_5 \widetilde{X}_A
+ i\epsilon_{\mu\nu\alpha\beta} n^\alpha \gamma^\beta\gamma_5 \widetilde{Y}_A
\nonumber\\&&{}\hspace*{0.1cm}
- (n_\mu v_\nu-n_\nu v_\mu)\slashed{n}\,{W} + (n_\mu \gamma_\nu-n_\nu \gamma_\mu)\slashed{n}\,{Z}
\biggr]\biggr\}(z_1,z_2;\mu)\,,
\label{def: 3-particle D-meson DAs}
\end{eqnarray}
where the convention corresponds $\epsilon_{0123}=-1$. Then one obtains three-particle higher-twist contributions to the correlation function of vacuum-to-$D$-meson at tree-level \cite{Lu:2018cfc}
\begin{eqnarray}
\Pi_{\mu, V}^{(3P)}(n \cdot p,\bar n \cdot p) &=&
-{\tilde{f}_D(\mu) \, m_D \over n \cdot p } \, \int_0^{\infty} \, d \omega_1  \, \int_0^{\infty} \, d \omega_2 \,
\int_0^1 d u \, {1 \over \left [\bar n \cdot p - \omega_1 - u \, \omega_2 \right ]^2} \nonumber \\
&&  \times \, \bigg \{ \bar n_{\mu} \, \left [ \rho_{\bar n, \rm{LP}}^{(3P)}(u, \omega_1, \omega_2, \mu)
+ {m_q \over n \cdot p} \, \rho_{\bar n, \rm{NLP}}^{(3P)}(u, \omega_1, \omega_2, \mu) \right ] \nonumber \\
&& +\,  n_{\mu} \, \left [ \rho_{n, \rm{LP}}^{(3P)}(u, \omega_1, \omega_2, \mu)
+ {m_q \over n \cdot p} \, \rho_{n, \rm{NLP}}^{(3P)}(u, \omega_1, \omega_2, \mu) \right ] \bigg \}   \,, \nonumber \\
\Pi_{\mu, T}^{(3P)}(n \cdot p,\bar n \cdot p) &=&  {i \over 2} \, {\tilde{f}_B(\mu) \, m_D^2 \over  n \cdot p } \,
\, \left [ n_{\mu} -  {n \cdot q \over m_D} \, \bar n_{\mu} \right ]
\, \int_0^{\infty} \, d \omega_1  \, \int_0^{\infty} \, d \omega_2 \,
\int_0^1 d u \, {1 \over \left [\bar n \cdot p - \omega_1 - u \, \omega_2 \right ]^2} \nonumber \\
&& \times  \bigg \{  \rho_{T, \rm{LP}}^{(3P)}(u, \omega_1, \omega_2, \mu)
+ {m_q \over n \cdot p} \, \rho_{T, \rm{NLP}}^{(3P)}(u, \omega_1, \omega_2, \mu)  \bigg \}  \,,
\end{eqnarray}
where the $m_q$ terms lead to the SU(3) flavor symmetry breaking effect, and the expressions of $\rho_{i, \rm{LP}}^{(3P)}$ and $\rho_{i, \rm{NLP}}^{(3P)}$ ($i=n\,, \bar n \,, T$) are collected in the appendix.

The two-particle higher-twist $D$-meson LCDAs are defined as \cite{Braun:2017liq}
\begin{eqnarray}
&& \langle  0 | \left (\bar d \, Y_s \right)_{\beta} (x) \,
\left (Y_s^{\dag} \, h_v \right )_{\alpha}(0)| \bar D(v)\rangle \nonumber \\
&& = - \frac{i \tilde f_D(\mu) \, m_D}{4}  \,
\int_0^{\infty} \, d \omega \, e^{- i \, \omega \, v \cdot x} \,
\bigg \{  \frac{1+ \! \not v}{2} \, \,
\bigg [ 2 \, \left ( \phi_{D}^{+}(\omega, \mu) + x^2 \, g_D^{+}(\omega, \mu)  \right ) \nonumber \\
&& \hspace{0.5 cm} - {1 \over v \cdot x}  \,
\left  [ \left ( \phi_{D}^{+}(\omega, \mu) - \phi_{D}^{-}(\omega, \mu)  \right )
+  x^2 \, \left ( g_{D}^{+}(\omega, \mu) - g_{D}^{-}(\omega, \mu)  \right )   \right ]  \,
\! \not x \bigg ] \, \gamma_5 \bigg \}_{\alpha \beta}\,,
\label{def: two-particle D-meson DAs with light-cone correction}
\end{eqnarray}
where $g_D^+$ and $g_D^-$ are of twist-four and twist-five, respectively. One could expand $g_D^+$ and $g_D^-$ in terms of three-particle LCDAs with the operator identities \cite{Kawamura:2001jm,Braun:2017liq}
\begin{align}
  \frac{\partial}{\partial x^\mu} \bar q(x)\gamma^\mu \Gamma [x,0] h_v(0)
&=
- i\int_0^1\! udu\, \bar q(x)[x,ux] x^\rho gG_{\rho\mu}(ux)[ux,0] \gamma^\mu \Gamma h_v(0)\,,
\notag\\
 v^\mu \frac{\partial}{\partial x^\mu} \bar q(x) \Gamma [x,0] h_v(0)
&=
 \phantom{-} i\int_0^1\! \bar udu\, \bar q(x)[x,ux] x^\rho gG_{\rho\mu}(ux)[ux,0] v^\mu \Gamma h_v(0)
\notag\\& \qquad{}
 + (v\cdot\partial) \bar q(x) \Gamma [x,0] h_v(0)\,.
\label{identity}
\end{align}
Applying the relations between two forms of $D$-meson LCDAs \cite{Lu:2018cfc}
\begin{eqnarray}
\Phi_3(\omega_1, \omega_2, \mu) &=&  \Psi_A(\omega_1, \omega_2, \mu) - \Psi_V(\omega_1, \omega_2, \mu) \,, \nonumber \\
\Phi_4(\omega_1, \omega_2, \mu) &=&  \Psi_A(\omega_1, \omega_2, \mu) + \Psi_V(\omega_1, \omega_2, \mu) \,, \nonumber \\
\Psi_4(\omega_1, \omega_2, \mu) &=&  \Psi_A(\omega_1, \omega_2, \mu) + X_A(\omega_1, \omega_2, \mu) \,, \nonumber \\
\tilde{\Psi}_4(\omega_1, \omega_2, \mu) &=&  \Psi_V(\omega_1, \omega_2, \mu) - \tilde{X}_A(\omega_1, \omega_2, \mu) \,, \nonumber \\
\Phi_5(\omega_1, \omega_2, \mu) &=&  \Psi_A(\omega_1, \omega_2, \mu) + \Psi_V(\omega_1, \omega_2, \mu)
+ 2 \, \left  [ Y_A -  \tilde{Y}_A +  W \right ] (\omega_1, \omega_2, \mu)\,, \nonumber \\
\Psi_5(\omega_1, \omega_2, \mu) &=&  - \Psi_A(\omega_1, \omega_2, \mu) + X_A(\omega_1, \omega_2, \mu)
- 2 \, Y_A(\omega_1, \omega_2, \mu) \,, \nonumber \\
\tilde{\Psi}_5(\omega_1, \omega_2, \mu) &=&  - \Psi_V(\omega_1, \omega_2, \mu) - \tilde{X}_A(\omega_1, \omega_2, \mu)
+ 2 \, \tilde{Y}_A(\omega_1, \omega_2, \mu) \,, \nonumber \\
\Phi_6(\omega_1, \omega_2, \mu) &=&  \Psi_A(\omega_1, \omega_2, \mu) - \Psi_V(\omega_1, \omega_2, \mu)
+ 2 \, \left  [ Y_A  +  \tilde{Y}_A
+  W - 2 \, Z \right ] (\omega_1, \omega_2, \mu)  \,,
\label{3P D-meson DAs of definite twist}
\end{eqnarray}
the nontrivial relations of $D$-meson LCDAs in momentum space could be obtained
\begin{eqnarray}
- \omega \, {d \over d \omega} \, \phi_D^{-}(\omega, \mu)
&=& \phi_D^{+}(\omega, \mu) - 2 \, \int_0^{\infty} \, {d \omega_2 \over \omega_2^2} \,
\Phi_3(\omega, \omega_2, \mu) + 2 \,  \int_0^{\omega} \, \, {d \omega_2 \over \omega_2^2} \,
\Phi_3(\omega-\omega_2, \omega_2, \mu)  \nonumber \\
&& + \, 2 \,  \int_0^{\omega} \, \, {d \omega_2 \over \omega_2} \,
{d \over d \omega} \, \Phi_3(\omega-\omega_2, \omega_2, \mu) \,,
\label{the first EOM} \\
-2 \, {d^2 \over d \omega^2} \, g_D^{+}(\omega, \mu) &=&
\left [ {3 \over 2} + (\omega - \bar \Lambda) \, {d \over d \omega}   \right ] \, \phi_D^{+}(\omega, \mu)
- {1 \over 2}  \, \phi_D^{-}(\omega, \mu)
+ \int_0^{\infty} \, {d \omega_2 \over \omega_2 } \, {d \over d \omega} \, \Psi_4(\omega, \omega_2, \mu) \nonumber \\
&& - \int_0^{\infty} \, {d \omega_2 \over \omega_2^2 } \, \Psi_4(\omega, \omega_2, \mu)
+ \int_0^{\omega} \, {d \omega_2 \over \omega_2^2 } \, \Psi_4(\omega-\omega_2, \omega_2, \mu) \,,
\label{the second EOM}  \\
-2 \, {d^2 \over d \omega^2} \, g_D^{-}(\omega, \mu) &=&
\left [ {3 \over 2} + (\omega - \bar \Lambda) \, {d \over d \omega}   \right ] \, \phi_D^{-}(\omega, \mu)
- {1 \over 2}  \, \phi_D^{+}(\omega, \mu)
+ \int_0^{\infty} \, {d \omega_2 \over \omega_2 } \, {d \over d \omega} \, \Psi_5(\omega, \omega_2, \mu) \nonumber \\
&& - \int_0^{\infty} \, {d \omega_2 \over \omega_2^2 } \, \Psi_5(\omega, \omega_2, \mu)
+ \int_0^{\omega} \, {d \omega_2 \over \omega_2^2 } \, \Psi_5(\omega-\omega_2, \omega_2, \mu) \,,
\label{the third EOM}  \\
\phi_D^{-}(\omega, \mu) &=&  \left (2 \, \bar \Lambda - \omega \right ) \, {d \phi_D^{+}(\omega, \mu) \over d \omega}
- 2 \, \int_0^{\infty} \, {d \omega_2 \over \omega_2^2 } \, \Phi_4(\omega, \omega_2, \mu)   \nonumber \\
&& + \,  2 \, \int_0^{\omega} \, {d \omega_2 \over \omega_2} \,
\left ( {d \over  d \, \omega_2} +   {d  \over d \, \omega}  \right ) \, \Phi_4(\omega-\omega_2, \omega_2, \mu)  \nonumber \\
&& + \,  2 \,  \int_0^{\omega} \, {d \omega_2 \over \omega_2} \, {d \over d \omega} \, \, \Psi_4(\omega-\omega_2, \omega_2, \mu)
 - \, 2 \,  \int_0^{\infty} \, {d \omega_2 \over \omega_2} \, {d \over d \omega} \, \, \Psi_4(\omega, \omega_2, \mu) \,,
\label{the fourth EOM}
\end{eqnarray}
which is consistent with Fourier transformed results from \cite{Braun:2017liq} ( see  appendix for  details).
By inserting two-particle $D$-meson LCDAs into the correlation function, one could obtain two-particle higher-twist corrections to the vacuum-to-$D$-meson correlation function at tree-level
\begin{eqnarray}
\Pi_{\mu, \, V}^{\rm 2PHT} &=& - 4 \, {\tilde{f}_D(\mu) \, m_D \over n \cdot p } \, \bar n_{\mu} \,
\bigg \{ - {1 \over 2} \, \int_0^{\infty} \, d \omega_1 \, \int_0^{\infty} \, d \omega_2 \, \int_0^1 d u \,
{\bar u \, \Psi_5(\omega_1, \omega_2, \mu) \over (\bar n \cdot p - \omega_1 - u \, \omega_2)^2}  \nonumber \\
&& + \int_0^{\infty} \, {d \omega \over (\bar n \cdot p -\omega)^2} \, \hat{g}_D^{-}(\omega, \mu)  \bigg \}  \,,  \nonumber \\
\Pi_{\mu, \, T}^{\rm 2PHT} &=&  2 \,  i \, {\tilde{f}_D(\mu) \, m_D^2 \over n \cdot p } \,
\, \left [ n_{\mu} -  {n \cdot q \over m_D} \, \bar n_{\mu} \right ] \,
\bigg \{ - {1 \over 2} \, \int_0^{\infty} \, d \omega_1 \, \int_0^{\infty} \, d \omega_2 \, \int_0^1 d u \,
{\bar u \, \Psi_5(\omega_1, \omega_2, \mu)  \over (\bar n \cdot p - \omega_1 - u \, \omega_2)^2} \,  \nonumber \\
&& + \int_0^{\infty} \, {d \omega \over (\bar n \cdot p -\omega)^2} \, \hat{g}_D^{-}(\omega, \mu)  \bigg \}  \,,
\end{eqnarray}
where $\hat{g}_D^{-}(\omega, \mu)$ is given by
\begin{eqnarray}
\hat{g}_D^{-}(\omega, \mu) =  {1 \over 4} \, \int_{\omega}^{\infty}  \, d \rho \,
\bigg \{ (\rho - \omega) \, \left [ \phi_D^{+}(\rho) -  \phi_D^{-}(\rho) \right ]
- 2 \, (\bar \Lambda - \rho)  \, \phi_D^{-}(\rho) \bigg \}  \,.
\label{def: gBminhat}
\end{eqnarray}

Collecting two-particle and three-particle contributions at tree-level together and matching the hadronic- and partonic-level predictions with the aid of dispersion relation,
one obtains the following expressions after Borel transformation
\begin{eqnarray}
&& {f_{P} \, n \cdot p \over 2} \,\, {\rm exp} \left [- {m_{P}^2 \over n \cdot p \,\, \omega_M} \right ] \,\,
\left [  f_{D \to P}^{+, \, \rm HT}(q^2)  + \frac{m_D} {n \cdot p} \, f_{D \to P}^{0, \, \rm HT}(q^2)  \right ] \, \nonumber \\
&& = - {\tilde{f}_D(\mu) \, m_D \over n \cdot p}  \,
\bigg \{ e^{-\omega_s/\omega_M} \, H_{\bar n, \rm LP}^{\rm 2PHT}(\omega_s, \mu)
+  \int_0^{\omega_s} \, d \omega^{\prime}  \, {1 \over \omega_M} \,
e^{-\omega^{\prime}/\omega_M}  \, H_{\bar n, \rm LP}^{\rm 2PHT}(\omega^{\prime}, \mu) \nonumber \\
&& \hspace{0.4 cm} + \int_0^{\omega_s} \, d \omega_1 \, \int_{\omega_s - \omega_1}^{\infty} \, {d \omega_2 \over \omega_2} \,
e^{-\omega_s/\omega_M} \,
\bigg [ H_{\bar n, \rm LP}^{\rm 3PHT} \left ({\omega_s - \omega_1 \over \omega_2}, \omega_1, \omega_2, \mu \right ) \nonumber \\
&& \hspace{0.8 cm} + {m_q \over n \cdot p} \,
H_{\bar n, \rm NLP}^{\rm 3PHT} \left ({\omega_s - \omega_1 \over \omega_2}, \omega_1, \omega_2, \mu \right ) \bigg ] \nonumber \\
&& \hspace{0.4 cm} + \int_0^{\omega_s} \, d \omega^{\prime} \, \int_0^{\omega^{\prime}} \, d \omega_1 \,
\int_{\omega^{\prime}  - \omega_1}^{\infty} \, {d \omega_2 \over \omega_2} \, {1 \over \omega_M} \, e^{-\omega^{\prime}/\omega_M} \,
\bigg [ H_{\bar n, \rm LP}^{\rm 3PHT} \left ({\omega^{\prime} - \omega_1 \over \omega_2}, \omega_1, \omega_2, \mu \right )
\nonumber \\
&& \hspace{0.8 cm} + {m_q \over n \cdot p} \,H_{\bar n, \rm NLP}^{\rm 3PHT}
\left ({\omega^{\prime} - \omega_1 \over \omega_2}, \omega_1, \omega_2, \mu \right ) \bigg ] \bigg \} \,,
\label{higher twist of fplus}
\\
&& {f_{P} \, n \cdot p \over 2} \,\, {\rm exp} \left [- {m_{P}^2 \over n \cdot p \,\, \omega_M} \right ] \,
{m_D \over n \cdot p -m_D} \,
\left [  f_{D \to P}^{+, \, \rm HT}(q^2)  - \frac{m_D} {n \cdot p} \, f_{D \to P}^{0, \, \rm HT}(q^2)  \right ] \, \nonumber \\
&& = - {\tilde{f}_D(\mu) \, m_D \over n \cdot p}  \,
\bigg \{  \int_0^{\omega_s} \, d \omega_1 \, \int_{\omega_s - \omega_1}^{\infty} \, {d \omega_2 \over \omega_2} \,
e^{-\omega_s/\omega_M} \,
\bigg [ H_{n, \rm LP}^{\rm 3PHT} \left ({\omega_s - \omega_1 \over \omega_2}, \omega_1, \omega_2, \mu \right ) \nonumber \\
&& \hspace{0.8 cm} + {m_q \over n \cdot p} \,
H_{n, \rm NLP}^{\rm 3PHT} \left ({\omega_s - \omega_1 \over \omega_2}, \omega_1, \omega_2, \mu \right ) \bigg ] \nonumber \\
&& \hspace{0.4 cm} + \int_0^{\omega_s} \, d \omega^{\prime} \, \int_0^{\omega^{\prime}} \, d \omega_1 \,
\int_{\omega^{\prime}  - \omega_1}^{\infty} \, {d \omega_2 \over \omega_2} \, {1 \over \omega_M} \, e^{-\omega^{\prime}/\omega_M} \,
\bigg [ H_{n, \rm LP}^{\rm 3PHT} \left ({\omega^{\prime} - \omega_1 \over \omega_2}, \omega_1, \omega_2, \mu \right )
\nonumber \\
&& \hspace{0.8 cm} + {m_q \over n \cdot p} \,H_{n, \rm NLP}^{\rm 3PHT}
\left ({\omega^{\prime} - \omega_1 \over \omega_2}, \omega_1, \omega_2, \mu \right ) \bigg ]  \bigg \} \,,
\label{higher twist of fzero}
\\
&& f_{P} \, n \cdot p \, {\rm exp} \left [- {m_{P}^2 \over n \cdot p \,\, \omega_M} \right ] \,\,
f_{D \to P}^{T, \, \rm HT}(q^2)  \, \nonumber \\
&& = - {\tilde{f}_D(\mu) \, (m_D +m_P) \over n \cdot p}  \,
\bigg \{ e^{-\omega_s/\omega_M} \, H_{T, \rm LP}^{\rm 2PHT}(\omega_s, \mu)
+  \int_0^{\omega_s} \, d \omega^{\prime}  \, {1 \over \omega_M} \,
e^{-\omega^{\prime}/\omega_M}  \, H_{T, \rm LP}^{\rm 2PHT}(\omega^{\prime}, \mu) \nonumber \\
&& \hspace{0.4 cm} + \int_0^{\omega_s} \, d \omega_1 \, \int_{\omega_s - \omega_1}^{\infty} \, {d \omega_2 \over \omega_2} \,
e^{-\omega_s/\omega_M} \,
\bigg [ H_{T, \rm LP}^{\rm 3PHT} \left ({\omega_s - \omega_1 \over \omega_2}, \omega_1, \omega_2, \mu \right ) \nonumber \\
&& \hspace{0.8 cm} + {m_q \over n \cdot p} \,
H_{T, \rm NLP}^{\rm 3PHT} \left ({\omega_s - \omega_1 \over \omega_2}, \omega_1, \omega_2, \mu \right ) \bigg ] \nonumber \\
&& \hspace{0.4 cm} + \int_0^{\omega_s} \, d \omega^{\prime} \, \int_0^{\omega^{\prime}} \, d \omega_1 \,
\int_{\omega^{\prime}  - \omega_1}^{\infty} \, {d \omega_2 \over \omega_2} \, {1 \over \omega_M} \, e^{-\omega^{\prime}/\omega_M} \,
\bigg [ H_{T, \rm LP}^{\rm 3PHT} \left ({\omega^{\prime} - \omega_1 \over \omega_2}, \omega_1, \omega_2, \mu \right )
\nonumber \\
&& \hspace{0.8 cm} + {m_q \over n \cdot p} \,H_{T, \rm NLP}^{\rm 3PHT}
\left ({\omega^{\prime} - \omega_1 \over \omega_2}, \omega_1, \omega_2, \mu \right ) \bigg ] \bigg \} \,,
\label{higher twist of fT}
\end{eqnarray}
where $H_{i, \rm LP}^{\rm 2PHT}$ and  $H_{i, \rm (N)LP}^{\rm 3PHT}$
($i = n, \, \bar n, \, T$) are given by
\begin{eqnarray}
H_{\bar n, \rm LP}^{\rm 2PHT}(\omega, \mu) &=&
 H_{T, \rm LP}^{\rm 2PHT}(\omega, \mu) = 4 \, \hat{g}_D^{-}(\omega, \mu) \,, \nonumber \\
 H_{n, \rm LP}^{\rm 3PHT} (u, \omega_1, \omega_2, \mu)&=& 2 \, (u-1) \, \Phi_4(\omega_1, \omega_2, \mu)  \,, \nonumber \\
H_{n, \rm NLP}^{\rm 3PHT} (u, \omega_1, \omega_2, \mu)&=& \tilde{\Psi}_5(\omega_1, \omega_2, \mu)
- \Psi_5(\omega_1, \omega_2, \mu)  \,, \nonumber \\
H_{\bar n, \rm LP}^{\rm 3PHT} (u, \omega_1, \omega_2, \mu)&=& \tilde{\Psi}_5(\omega_1, \omega_2, \mu)
- \Psi_5(\omega_1, \omega_2, \mu)  \,, \nonumber \\
H_{\bar n, \rm NLP}^{\rm 3PHT} (u, \omega_1, \omega_2, \mu)&=&  2 \,\Phi_6(\omega_1, \omega_2, \mu)  \,, \nonumber \\
H_{T, \rm LP}^{\rm 3PHT} (u, \omega_1, \omega_2, \mu)&=& 2\, (1-u) \,\Phi_4(\omega_1, \omega_2, \mu)
- \Psi_5(\omega_1, \omega_2, \mu)  + \tilde{\Psi}_5(\omega_1, \omega_2, \mu)  \,, \nonumber \\
H_{T, \rm NLP}^{\rm 3PHT} (u, \omega_1, \omega_2, \mu)&=&  \Psi_5(\omega_1, \omega_2, \mu)
- \tilde{\Psi}_5(\omega_1, \omega_2, \mu)  + 2 \, \Phi_6(\omega_1, \omega_2, \mu)    \,.
\end{eqnarray}
One can observe from the  equations (\ref{higher twist of fplus})-(\ref{higher twist of fT}) that the large recoil symmetry is broken by three-particle higher-twist contributions.
Finally, we obtain the sum rules for $D\to P$ form factors as follows
\begin{eqnarray}
f_{D \to P}^{+}(q^2) &=&  f_{D \to P}^{+, \rm 2PNLL}(q^2) + f_{D \to P}^{+, \, \rm 2PHT}(q^2)
+ f_{D \to P}^{+, \, \rm 3PHT}(q^2)  \,, \nonumber \\
f_{D \to P}^{0}(q^2) &=&  f_{D \to P}^{0, \rm 2PNLL}(q^2) + f_{D \to P}^{0, \, \rm 2PHT}(q^2)
+ f_{D \to P}^{0, \, \rm 3PHT}(q^2)  \,, \nonumber  \\
f_{D \to P}^{T}(q^2) &=&  f_{D \to P}^{T, \rm 2PNLL}(q^2) + f_{D \to P}^{T, \, \rm 2PHT}(q^2)
+ f_{D \to P}^{T, \, \rm 3PHT}(q^2)  \,
\label{final sum rules}
\end{eqnarray}
where all the three form factors are of $\mathcal{ O}(\Lambda/m_c)^{5/2}$.

\section{Numerical Analysis}

\subsection{Models of the higher-twist $D$-meson LCDAs}
$D$-meson LCDAs are the fundamental nonperturbative inputs in the $D$-meson LCSR. The model of the higher-twist $D$-meson LCDAs can be expressed with the matrix elements of local operators \cite{Grozin:1996pq}
\begin{eqnarray}
\langle 0 | \bar q \, g_s \, G_{\mu \nu} \, \Gamma \, h_v | \bar D(v) \rangle
&=& - {\tilde{f}_D(\mu) \, m_D \over 6} \,
\bigg \{ i \, \lambda_H^2 {\rm Tr} \left [ \gamma_5 \, \Gamma \, {1 + \not \! v \over 2}  \, \sigma_{\mu \nu} \right ] \nonumber \\
&& + \,  (\lambda_H^2 - \lambda_E^2)  \,{\rm Tr}  \left [ \gamma_5 \, \Gamma \, {1 + \not \! v \over 2}  \,
(v_{\mu} \, \gamma_{\nu} - v_{\nu} \, \gamma_{\mu} ) \right ]  \bigg  \}  \,.
\label{HQET parametrization of the 3P matrix element}
\end{eqnarray}
Implementing the standard strategy of LCSR, one obtain the three-particle higher-twist $D$-meson LCDA sum rules as follows \cite{Lu:2018cfc}
\begin{eqnarray}
&& [\tilde{f}_D(\mu)]^2  \, m_D \, (\lambda_H^2 + \lambda_E^2) \,
\Phi_5(\omega_1, \omega_2, \mu) \,\nonumber \\
&&  = - {g_s^2 \, C_F \, N_c \over 96 \, \pi^4} \,
\int_{\omega_1 + \omega_2 \over 2}^{\omega_0} \, ds \,
{\rm exp} \left [{\bar \Lambda - s \over \omega_M}  \right ]  \,
\omega_1 \, (\omega_1 + \omega_2 - 2 \, s)^3 \, \theta(2 \, s - \omega_1 - \omega_2)   \,,
\nonumber \\
&& [\tilde{f}_D(\mu)]^2  \, m_D \, (\lambda_H^2 + \lambda_E^2) \,
\Psi_5(\omega_1, \omega_2, \mu) \,\nonumber \\
&&  =  {g_s^2 \, C_F \, N_c \over 192 \, \pi^4} \,
\int_{\omega_1 + \omega_2 \over 2}^{\omega_0} \, ds \,
{\rm exp} \left [{\bar \Lambda - s \over \omega_M}  \right ]  \,
\omega_2 \, (\omega_1 + \omega_2 - 2 \, s)^3 \, \theta(2 \, s - \omega_1 - \omega_2)   \,,
\nonumber \\
&& [\tilde{f}_D(\mu)]^2  \, m_D \, (\lambda_H^2 + \lambda_E^2) \,
\tilde{\Psi}_5(\omega_1, \omega_2, \mu) \,\nonumber \\
&&  =  {g_s^2 \, C_F \, N_c \over 192 \, \pi^4} \,
\int_{\omega_1 + \omega_2 \over 2}^{\omega_0} \, ds \,
{\rm exp} \left [{\bar \Lambda - s \over \omega_M}  \right ]  \,
\omega_2 \, (\omega_1 + \omega_2 - 2 \, s)^3 \, \theta(2 \, s - \omega_1 - \omega_2)   \,,
\nonumber \\
&& [\tilde{f}_D(\mu)]^2  \, m_D \, (\lambda_E^2 - \lambda_H^2) \,
\Phi_6(\omega_1, \omega_2, \mu) \,\nonumber \\
&&  =  {g_s^2 \, C_F \, N_c \over 128 \, \pi^4} \,
\int_{\omega_1 + \omega_2 \over 2}^{\omega_0} \, ds \,
{\rm exp} \left [{\bar \Lambda - s \over \omega_M}  \right ]  \,
(\omega_1 + \omega_2 - 2 \, s)^4 \, \theta(2 \, s - \omega_1 - \omega_2)   \,,
\label{QCDSR for 3P B-meson DAs}
\end{eqnarray}

We first introduce the local duality model by taking the limit $\omega_M\to\infty$ of the sum rules (\ref{QCDSR for 3P B-meson DAs}).
Using the normalization conditions \cite{Lu:2018cfc}
\begin{eqnarray}
\Phi_5(z_1=z_2=0, \mu) &=& \int_0^{\infty} \, d \omega_1 \, \int_0^{\infty} \, d \omega_2 \,\, \Phi_5(\omega_1, \omega_2, \mu)
= {\lambda_E^2 + \lambda_H^2 \over 3} \,,  \nonumber \\
\Psi_5(z_1=z_2=0, \mu) &=& \int_0^{\infty} \, d \omega_1 \, \int_0^{\infty} \, d \omega_2 \,\, \Psi_5(\omega_1, \omega_2, \mu)
= - {\lambda_E^2 \over 3} \,,  \nonumber \\
\tilde{\Psi}_5(z_1=z_2=0, \mu) &=& \int_0^{\infty} \, d \omega_1 \, \int_0^{\infty} \, d \omega_2 \,\, \tilde{\Psi}_5(\omega_1, \omega_2, \mu)
= - {\lambda_H^2 \over 3} \,,  \nonumber \\
\Phi_6(z_1=z_2=0, \mu) &=& \int_0^{\infty} \, d \omega_1 \, \int_0^{\infty} \, d \omega_2 \,\, \Phi_6(\omega_1, \omega_2, \mu)
=  {\lambda_E^2 - \lambda_H^2 \over 3} \,,
\label{normalization conditions of twist 5 and 6 DAs}
\end{eqnarray}
one could obtain the local duality models for twist-five and twist-six $D$-meson LCDAs
\begin{eqnarray}
\Phi_5^{\rm LD}(\omega_1, \omega_2, \mu)  &=& {35 \over 64} \, (\lambda_E^2 + \lambda_H^2) \,
{\omega_1 \over \omega_0^7} \, (2 \, \omega_0 - \omega_1-\omega_2)^4 \,
\theta(2 \, \omega_0 - \omega_1 - \omega_2)  \,, \nonumber \\
\Psi_5^{\rm LD}(\omega_1, \omega_2, \mu)  &=& - {35 \over 64} \, \lambda_E^2 \,
{\omega_2 \over \omega_0^7} \, (2 \, \omega_0 - \omega_1-\omega_2)^4 \,
\theta(2 \, \omega_0 - \omega_1 - \omega_2)  \,, \nonumber \\
\tilde{\Psi}_5^{\rm LD}(\omega_1, \omega_2, \mu)  &=& - {35 \over 64} \, \lambda_H^2 \,
{\omega_2 \over \omega_0^7} \, (2 \, \omega_0 - \omega_1-\omega_2)^4 \,
\theta(2 \, \omega_0 - \omega_1 - \omega_2)  \,, \nonumber \\
\Phi_6^{\rm LD}(\omega_1, \omega_2, \mu)  &=& {7 \over 64} \, (\lambda_E^2 - \lambda_H^2) \,
{1 \over \omega_0^7} \, (2 \, \omega_0 - \omega_1-\omega_2)^5 \,
\theta(2 \, \omega_0 - \omega_1 - \omega_2)  \,,
\label{LD model for the 3P D-meson DAs}
\end{eqnarray}
and they satisfy the following asymptotic behaviours \cite{Lu:2018cfc}
\begin{eqnarray}
\Phi_5(\omega_1, \omega_2, \mu) \sim \omega_1 \,, \qquad
\Psi_5(\omega_1, \omega_2, \mu)  \sim
\tilde{\Psi}_5(\omega_1, \omega_2, \mu) \sim \omega_2 \,,
\qquad
\Phi_6(\omega_1, \omega_2, \mu) \sim 1 \,.
\label{asymtotic behaviour of twist 5 and 6 DAs}
\end{eqnarray}
The remaining models for two-particle and three-particle $D$-meson LCDAs are given by \cite{Braun:2017liq}
\begin{eqnarray}
\phi_D^{+, \rm LD}(\omega, \mu) &=& {5 \over 8 \, \omega_0^5}  \, \omega(2 \, \omega_0 - \omega)^3 \,
\theta(2 \, \omega_0 - \omega)\,,  \nonumber  \\
\phi_D^{-, \rm LD}(\omega, \mu) &=& {5 (2 \, \omega_0 - \omega)^2 \over 192\, \omega_0^5}  \,
\bigg \{6 \, (2 \, \omega_0 - \omega)^2 - {7 \, (\lambda_E^2 - \lambda_H^2) \over \omega_0^2} \,
(15 \, \omega^2 - 20 \, \omega \, \omega_0 + 4 \, \omega_0^2) \bigg \}   \, \nonumber \\
&& \times \, \theta(2 \, \omega_0 - \omega)\,,  \nonumber  \\
\Phi_3^{\rm LD}(\omega_1, \omega_2, \mu) &=&  {105 \, (\lambda_E^2 - \lambda_H^2) \over 8 \, \omega_0^7} \,
\omega_1 \, \omega_2^2 \, \left (\omega_0 - {\omega_1 + \omega_2 \over 2} \right )^2  \,
\theta(2 \, \omega_0 - \omega_1 - \omega_2)  \,,  \nonumber  \\
\Phi_4^{\rm LD}(\omega_1, \omega_2, \mu) &=&  {35 \, (\lambda_E^2 + \lambda_H^2) \over 4 \, \omega_0^7} \,
\omega_2^2 \, \left (\omega_0 - {\omega_1 + \omega_2 \over 2} \right )^3  \,
\theta(2 \, \omega_0 - \omega_1 - \omega_2)  \,,  \nonumber  \\
\Psi_4^{\rm LD}(\omega_1, \omega_2, \mu) &=&  {35 \, \lambda_E^2 \over 2 \, \omega_0^7} \,
\omega_1 \, \omega_2 \, \left (\omega_0 - {\omega_1 + \omega_2 \over 2} \right )^3  \,
\theta(2 \, \omega_0 - \omega_1 - \omega_2)  \,,  \nonumber  \\
\tilde{\Psi}_4^{\rm LD}(\omega_1, \omega_2, \mu) &=&  {35 \, \lambda_H^2 \over 2 \, \omega_0^7} \,
\omega_1 \, \omega_2 \, \left (\omega_0 - {\omega_1 + \omega_2 \over 2} \right )^3  \,
\theta(2 \, \omega_0 - \omega_1 - \omega_2)  \,.
\end{eqnarray}
The effective LCDA defined in (\ref{def: gBminhat}) can be obtained using the above formulae
\begin{eqnarray}
\hat{g}_D^{-, \rm LD}(\omega, \mu) &=& {\omega \, (2 \, \omega_0 - \omega)^3 \over \omega_0^5} \,
\left \{ {5 \over 256} \, (2 \, \omega_0 - \omega)^2  - {35 \, (\lambda_E^2 - \lambda_H^2) \over 1536} \,
\left [ 4 - 12 \, \left ( {\omega \over \omega_0} \right )
+ 11 \, \left ( {\omega \over \omega_0} \right )^2  \right ]\right \}  \nonumber \\
&& \times \, \theta(2 \, \omega_0 - \omega) \,,
\end{eqnarray}
with the equation of motion (EOM) constraint \cite{Braun:2017liq}
\begin{eqnarray}
\omega_0 = {5 \over 2} \, \lambda_D = 2 \, \bar \Lambda\,, \qquad
3 \, \omega_0^2 = 14 \, (2 \, \lambda_E^2 + \lambda_H^2)\,.
\label{HQET relations for LD model}
\end{eqnarray}
Combining with the asymptotic behaviours (\ref{asymtotic behaviour of twist 5 and 6 DAs}),
the exponential model for twist-five and twist-six $D$-meson LCDAs could be obtained by implementing an exponential falloff at large momenta
\begin{eqnarray}
\Phi_5^{\rm exp}(\omega_1, \omega_2, \mu)
&=& {\lambda_E^2 + \lambda_H^2 \over 3 \, \omega_0^3} \, \omega_1 \,
e^{-(\omega_1 + \omega_2)/\omega_0} \,, \nonumber \\
\Psi_5^{\rm exp}(\omega_1, \omega_2, \mu)
&=& - {\lambda_E^2 \over 3 \, \omega_0^3} \, \omega_2 \,
e^{-(\omega_1 + \omega_2)/\omega_0} \,, \nonumber \\
\tilde{\Psi}_5^{\rm exp}(\omega_1, \omega_2, \mu)
&=& - {\lambda_H^2 \over 3 \, \omega_0^3} \, \omega_2 \,
e^{-(\omega_1 + \omega_2)/\omega_0} \,, \nonumber \\
\Phi_6^{\rm exp}(\omega_1, \omega_2, \mu)
&=& {\lambda_E^2 - \lambda_H^2 \over 3 \, \omega_0^2} \,
e^{-(\omega_1 + \omega_2)/\omega_0} \,.
\end{eqnarray}
Exponential models for the remaining $D$-meson LCDAs have been evaluated in \cite{Braun:2017liq}, and expressions of them are given by
\begin{eqnarray}
\phi_D^{+, \, \rm exp}(\omega, \mu) &=& {\omega \over \omega_0^2} \,  e^{-\omega/\omega_0} \,, \nonumber \\
\phi_D^{-, \, \rm exp}(\omega, \mu) &=& {1 \over \omega_0} \,  e^{-\omega/\omega_0}
- {\lambda_E^2 - \lambda_H^2 \over 9 \, \omega_0^3}  \,
\left [ 1 - 2 \, \left ( {\omega \over \omega_0} \right )
+ {1 \over 2} \, \left ( {\omega \over \omega_0} \right )^2 \right ]
\,  e^{-\omega/\omega_0} \,, \nonumber \\
\Phi_3^{\rm exp}(\omega_1, \omega_2, \mu) &=&
{\lambda_E^2 - \lambda_H^2 \over 6 \, \omega_0^5} \, \omega_1 \, \omega_2^2 \,
e^{-(\omega_1 + \omega_2)/\omega_0} \,, \nonumber \\
\Phi_4^{\rm exp}(\omega_1, \omega_2, \mu) &=&
{\lambda_E^2 + \lambda_H^2 \over 6 \, \omega_0^4} \, \omega_2^2 \,
e^{-(\omega_1 + \omega_2)/\omega_0} \,, \nonumber \\
\Psi_4^{\rm exp}(\omega_1, \omega_2, \mu) &=&
{\lambda_E^2 \over 3 \, \omega_0^4} \, \omega_1 \, \omega_2 \,
e^{-(\omega_1 + \omega_2)/\omega_0} \,, \nonumber \\
\tilde{\Psi}_4^{\rm exp}(\omega_1, \omega_2, \mu) &=&
{\lambda_H^2 \over 3 \, \omega_0^4} \, \omega_1 \, \omega_2 \,
e^{-(\omega_1 + \omega_2)/\omega_0} \,
\end{eqnarray}
and the expression of the two-particle twist-five LCDA reads
\begin{eqnarray}
\hat{g}_D^{-,  \, \rm exp}(\omega, \mu) = \omega \,
\left \{ {3 \over 4}  - {\lambda_E^2 - \lambda_H^2 \over 12 \, \omega_0^2} \,
\left [ 1 - \left ( {\omega \over \omega_0} \right )
+ {1 \over 3} \, \left ( {\omega \over \omega_0} \right )^2 \right ] \right \}
\,  e^{-\omega/\omega_0}  \,
\end{eqnarray}
with the EOM constrains \cite{Braun:2017liq}
\begin{eqnarray}
\omega_0 =  \lambda_D = {2 \over 3} \, \bar \Lambda\,, \qquad
2 \, \bar \Lambda^2 = 2 \, \lambda_E^2 + \lambda_H^2 \,.
\label{HQET relations for exponential model}
\end{eqnarray}

\begin{table}[htb]
\begin{tabular}{p{2cm}|p{4.2cm}|p{2cm}|p{4.2cm}}
\hline
Parameter          & DATA                                & Parameter                  & DATA \tabularnewline
\hline
$m_D $             & $1.86965\pm0.05\, \rm GeV $         & $\mu_{h1}$                 & $1.288\pm0.020\, \rm GeV $    \tabularnewline
$\tau_D$           & $(1.040\pm0.007)\times10^{-12}\,s$  & $\mu_{h2}$                 & $1.288\pm0.020\, \rm GeV $    \tabularnewline
$m_c$              & $1.288\pm0.020\, \rm GeV$           & $\mu_0$                    & $1\, \rm GeV             $    \tabularnewline
$m_d$              & $4.71\pm0.09\, \rm MeV $            & $f_D$                      & $212.0\pm0.7\, \rm MeV   $    \tabularnewline
$\lambda_D(\mu_0)$ & $0.354_{-30}^{+38}\,\rm  GeV$       & $\lambda_E^2/\lambda_H^2$  & $0.5\pm0.1\,\rm GeV^2    $    \tabularnewline
$\sigma_1(\mu_0)$  & $1.5\pm1$                           & $2\lambda_E^2+\lambda_H^2$ & $0.25\pm0.15\,\rm GeV^2  $    \tabularnewline
$\sigma_2(\mu_0)$  & $3\pm2$                             & $\bar{\Lambda}$            & $0.58\, \rm GeV          $    \tabularnewline
\hline
\end{tabular}
\centering{}\caption{ Parameters employed in our calculation, where the particle parameters from \cite{PDG:2020} and the others from \cite{Lu:2018cfc}. }
\label{values}
\end{table}

Two different $D$-meson LCDA models, the exponential model and the local duality model, are adopted in this work.
As three HQET parameters $\lambda_D(\mu)$, $\lambda_E(\mu)$ and $\lambda_H(\mu)$ are constrained by the EOM, and the ratio $R(\mu)$ ($R(\mu)=\lambda_E^2(\mu)/\lambda_H^2(\mu)$) is insensitive to perturbative and nonperturbative QCD corrections, we will take $R(\mu)$ and $\lambda_D(\mu)$ as the input in our numerical analysis. The renormalization scale dependence of the inverse moment is evaluated from the one loop equation of $\phi_D(\omega,\mu)$ \cite{Lange:2003aaa,Braun:2004aaa}
\begin{eqnarray}
\frac{\lambda_D(\mu_0)}{\lambda_D(\mu)} &=&
1 + {\alpha_s(\mu_0) \, C_F \over 4 \, \pi} \, \ln {\mu \over \mu_0} \,
\left [2 - 2\, \ln {\mu \over \mu_0} - 4 \, \sigma_{1}(\mu_0) \right ] + {\cal O}(\alpha_s^2)\,,
\label{lambdab evolution}
\end{eqnarray}
and the definition of the inverse-logarithmic moment $\sigma_{i}(\mu_0)$ is \cite{Beneke:2011nf}
\begin{equation}\label{inv-log}
  \sigma_n(\mu)=\lambda_D(\mu)\int_{0}^{\infty}\frac{d\omega}{\omega}{\rm ln}^n\frac{\mu_0}{\omega}\phi_{D}^+(\omega,\mu)
\end{equation}
We obtain the values of $\lambda_D(\mu_0)$ by matching our predicted zero momentum transfer $D\to\pi$ vector form factor with results from pion LCDAs $f_{D\to\pi}^+(0)=0.67_{-0.07}^{+0.10}$ \cite{Khodjamirian:2009ys}
\begin{eqnarray}
\lambda_D(\mu_0) = \left\{
\begin{array}{l}
260^{+24}_{-34} \,\, {\rm MeV}  \,, \qquad  \hspace{1.5 cm}
(\rm Exponential \,\, Model) \vspace{1.0 cm} \\
295^{+25}_{-33} \,\, {\rm MeV}  \,.
 \qquad  \hspace{1.5 cm}
(\rm Local \,\, Duality \,\, Model)
\end{array}
 \hspace{0.5 cm} \right.
\,
\label{values of lambdaD}
\end{eqnarray}

\subsection{Predicted form factors}
Except for the $D$-meson LCDAs, we present various values of input parameters in table \ref{values}. The mass and lifetime of hadrons, and quark masses in the $\overline{\text{MS}}$ scheme are taken from the Particle Data Group (PDG) \cite{PDG:2020}. The factorization scale interval $\mu\in[1,1.4]\,\rm GeV$ with the central value $1.2\,\rm GeV$ is the same as \cite{Lu:2021ttf}, where the maximum of the factorization scale is consistent with \cite{Khodjamirian:2009ys}. In the sum rules, we use the same internal sum rule parameters in \cite{Wang:2015vgv}
\begin{eqnarray}
M^2 = (1.25 \pm 0.25) \, {\rm GeV^2},\,
 s_0^{\pi}  = (0.70 \pm 0.05) \, {\rm GeV^2},\,
s_0^{K} = (1.05 \pm 0.05) \, {\rm GeV^2}.
\end{eqnarray}

\begin{figure}
  \centering
  \includegraphics[width=0.32\textwidth]{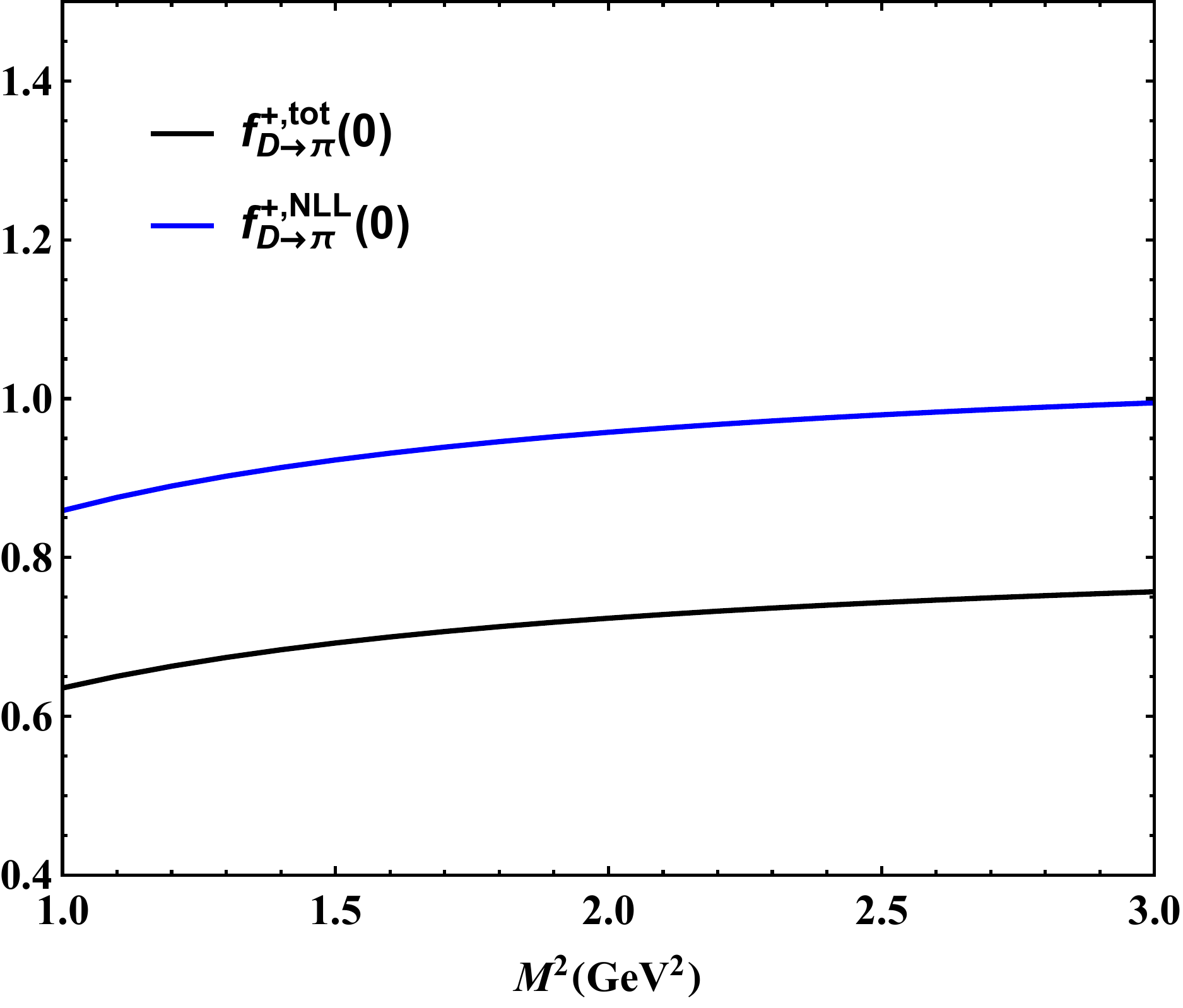}
  \includegraphics[width=0.32\textwidth]{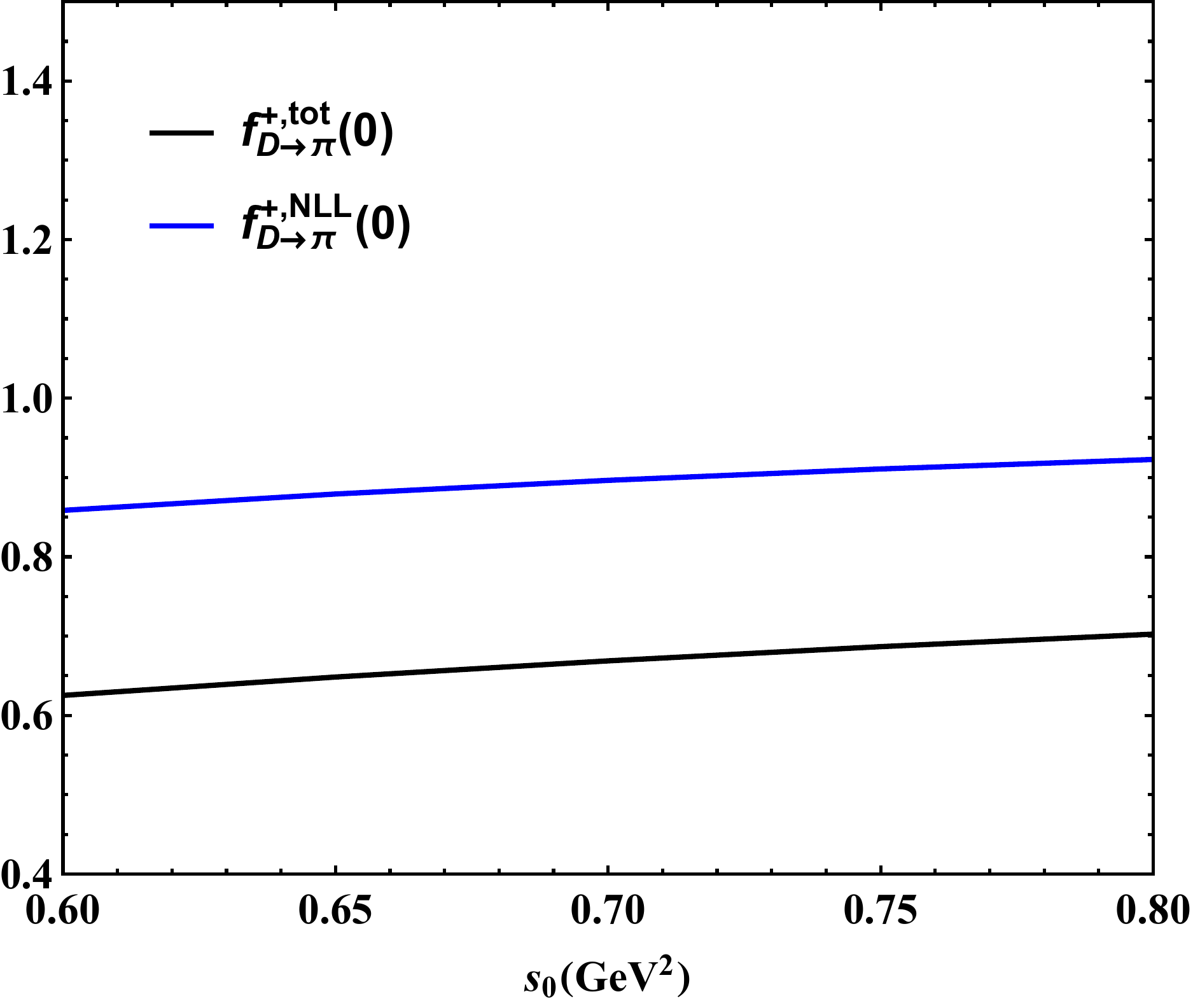}
  \includegraphics[width=0.32\textwidth]{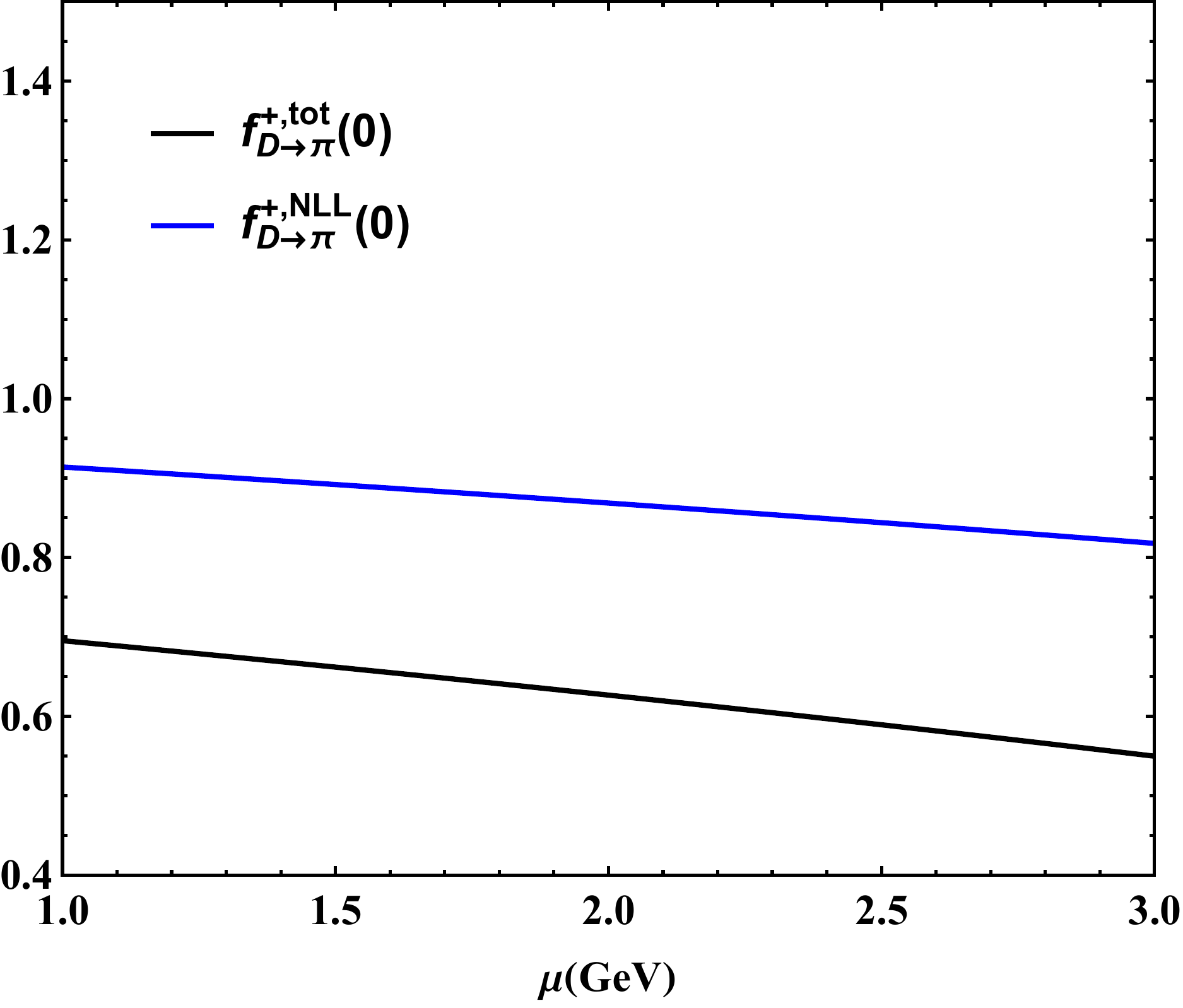}
  \caption{Dependencies of the vector $D\to\pi$ form factors on the Borel parameter $M^2$, the effective threshold $s_0$, and the factorization scale $\mu$.}\label{borel depend}
\end{figure}

\begin{figure}
  \centering
  \includegraphics[width=0.5\textwidth]{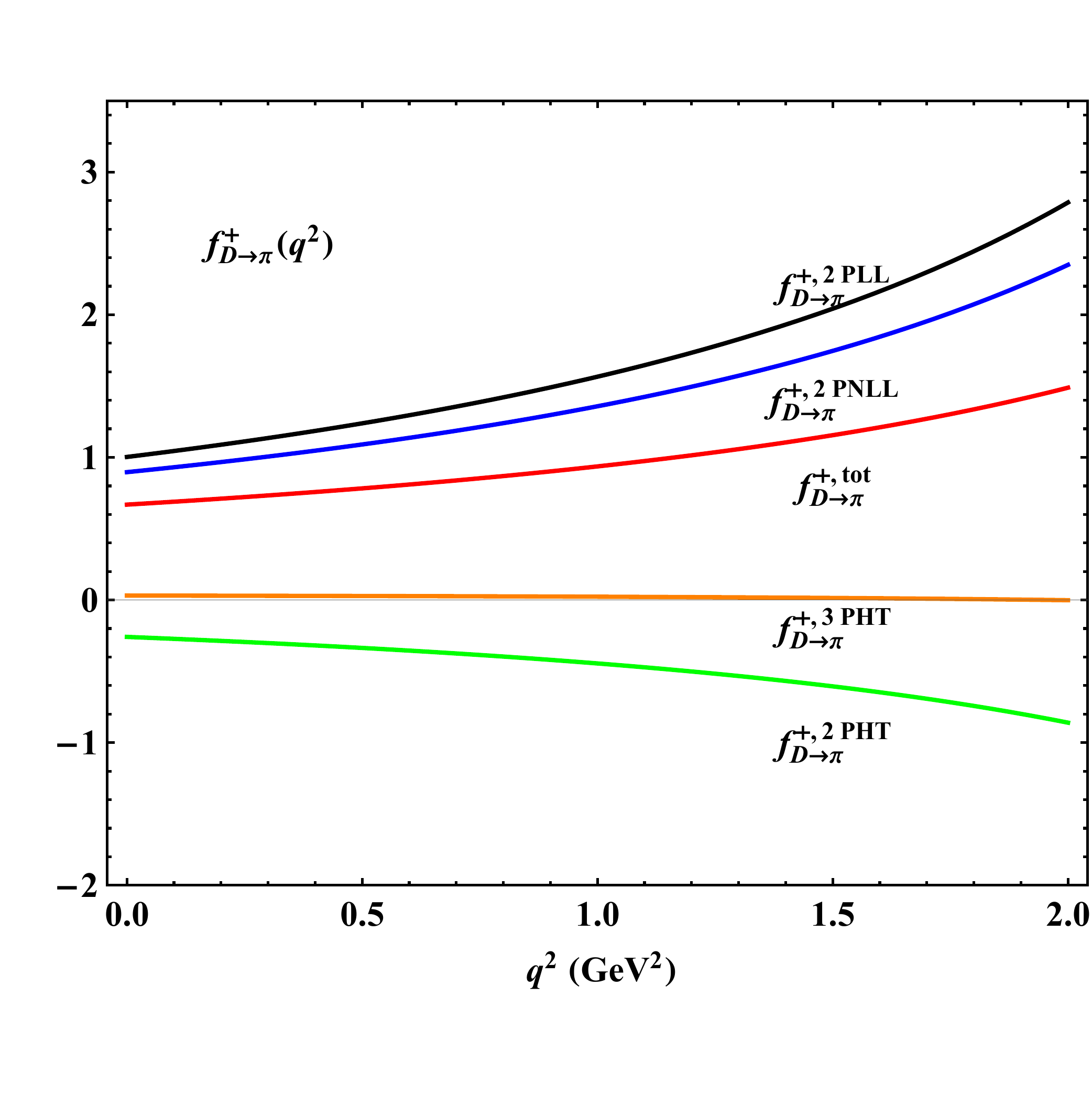}
  \caption{The momentum-transfer dependence of the vector $D\to\pi$ form factors, different form factors from the leading power LL contribution, the leading power NLL contribution, the two-particle higher-twist contribution and the three-particle higher-twist contribution are listed.}\label{figure1}
\end{figure}

\begin{figure}
  \centering
  \includegraphics[width=0.45\textwidth]{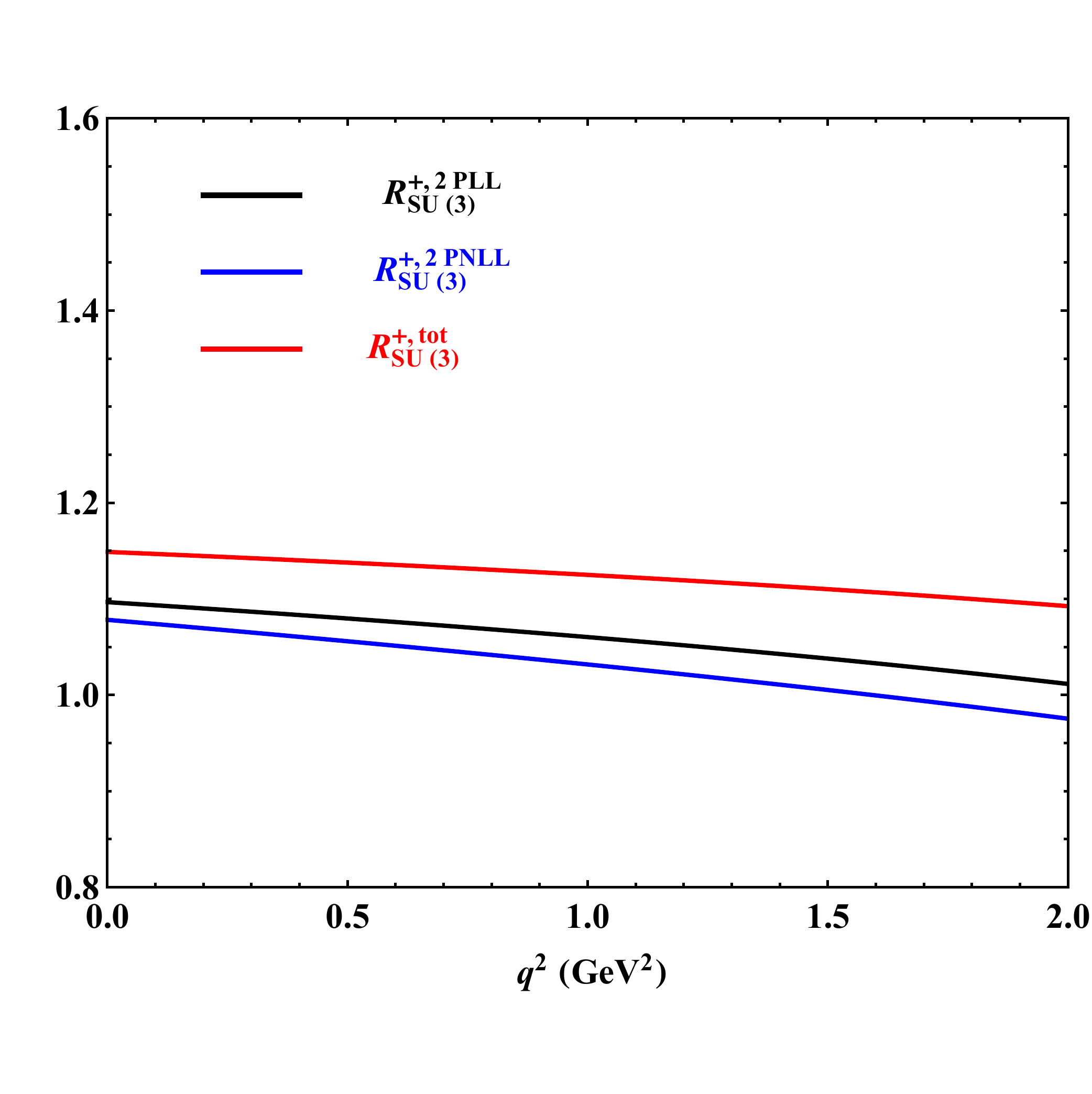}
  \includegraphics[width=0.45\textwidth]{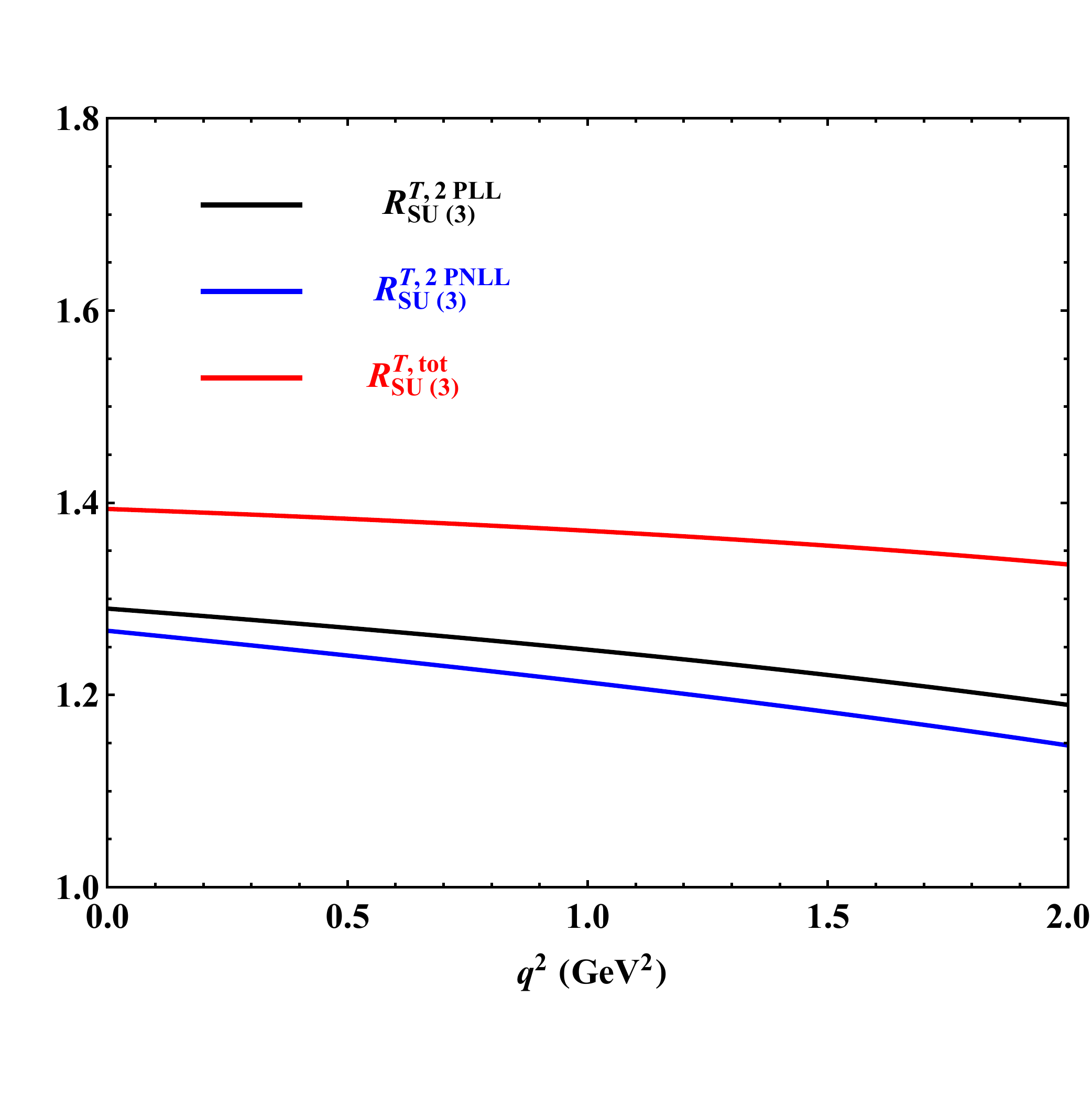}
  \caption{The SU(3)-flavor symmetry breaking effects between $D\to\pi$ and $D\to K$ form factors from the vector and tensor $c\to q$ weak current.}\label{figure2and3}
\end{figure}

In the following discussion, we will take the exponential model as the default model of $D$-meson LCDAs. We firstly focus on the breakdown of different contributions to the $D\to\pi$ vector form factors. As shown in Fig. \ref{figure1}, for the leading power contribution, the NLL resummation will lead to a $(5\%\sim17\%)$ reduction to the leading logarithmic (LL) result.  As for the power suppressed contributions, the two-particle twist-five contribution is the dominant contribution, which leads to correction of ${\mathcal O} (27\%\sim36\%)$ to the leading power form factor $f_{D\to\pi}^+(q^2)$ of $D\to\pi$. However, the three-particle higher-twist contribution is tiny, which reduces the form factor $f_{D\to\pi}^{+,\rm 2\,PNLL}$ of ${\cal O}(1\%)$.
The SU(3) flavor symmetry breaking effects between different final states of pion and kaon are defined as
\begin{eqnarray}
R_{\rm SU(3)}^{i} (q^2) =  \frac{f_{D \to K}^{i}(q^2)}{f_{D \to \pi}^{i}(q^2)} \,, \qquad
({\rm with}  \,\, i=+, \, 0, \, T) \,
\end{eqnarray}
and the results are presented in figure \ref{figure2and3}.
The SU(3) flavor symmetry breaking effects of the scalar and vector form factors $R_{\rm SU(3)}^{0,+} (0)$ are
adjusted to reproduce the results from pion LCDA \cite{Khodjamirian:2009ys},
and the tensor result $R_{\rm SU(3)}^{T} (0)$=1.39 agrees with the result from Lattice QCD \cite{Lubicz:2018rfs} $R_{\rm SU(3)}^{T} (0)$=1.36.
As we neglect the masses of the up and down quarks, the SU(3) flavor symmetry breaking effects originate from the strange quark mass, the difference between the pion and kaon threshold parameters, and the discrepancy between the decay constants $f_{\pi}$ and $f_{K}$. Different from the $B$-meson decays, the strange quark appears in the charged current of $D$ decays. The dependencies of the leading power form factor at NLL $f_{D\to\pi}^{+,\rm NLL}$ and total form factor $f_{D\to\pi}^{+,\rm tot}$ of $D\to\pi$ at $q^2=0$ on the Borel parameter $M^2$, the effective threshold parameter $s_0$, and the factorization scale $\mu$ are shown in Fig. \ref{borel depend}. Observing the left two panels, we find that the uncertainties from $M^2$ and $s_0$ are both of ${\cal O}(10\%)$, and the higher-twist contributions to the leading power form factor are insensitive to the HQET parameters.
As shown in the right panel of Fig. \ref{borel depend}, the uncertainty from the factorization scale is small in the interval of $\mu\in[1,1.4]\,\rm GeV$.

 To guarantee the reliability of the LCSR, we directly calculate the form factors at the space-like region  $q^2\in[-2,\,0]\,\rm GeV^2$. The results can be analytically continued to the positive kinematic region by applying the conformal transformation
\begin{eqnarray}
z(q^2, t_0) = \frac{\sqrt{t_{+}-q^2}-\sqrt{t_{+}-t_0}}
{\sqrt{t_{+}-q^2}+\sqrt{t_{+}-t_0}}  \,,
\end{eqnarray}
then the branching cut region of $q^2$ is mapped onto a disk $|z(q^2, \, t_0)|\leq 1$.
In the above equation, $t_{+}$ and $t_0$ are give by
\begin{eqnarray}
t_{+} &=& (m_B + m_{P})^2 \\
t_0   &=& (m_B + m_P) \, (\sqrt{m_B} + \sqrt{m_P})^2 \,.
\end{eqnarray}
The series expansion for $D\to P$ form factors is similar to the $B$-meson decays \cite{Bourrely:2008za} due to the heavy quark symmetry
\begin{eqnarray}
f_{D \to P}^{+, T}(q^2) &=&  {f_{D \to P}^{+, T}(0) \over 1 - q^2/m_{D_{(s)}^{\ast}}^2} \,
\bigg \{ 1 + \, \sum_{k=1}^{N-1}   \, b_{k, P}^{+, T}  \,
\bigg  ( z(q^2, \, t_0)^k -  z(0, \, t_0)^k  \nonumber \\
&& - \, (-1)^{N-k} \, {k \over N} \,
\left [  z(q^2, \, t_0)^N -  z(0, \, t_0)^N \right ]  \bigg  ) \bigg \}   \,,
\nonumber \\
f_{D \to P}^{0}(q^2) &=&  f_{D \to P}^{0}(0) \,
\left \{ 1 +  \, \sum_{k=1}^{N}   \, b_{k, P}^{0}  \,
\left (  z(q^2, \, t_0)^k -  z(0, \, t_0)^k  \right )   \right \}   \,,
\label{z expansion of B to P FFs}
\end{eqnarray}

\begin{figure}
  \centering
  \includegraphics[width=0.45\textwidth]{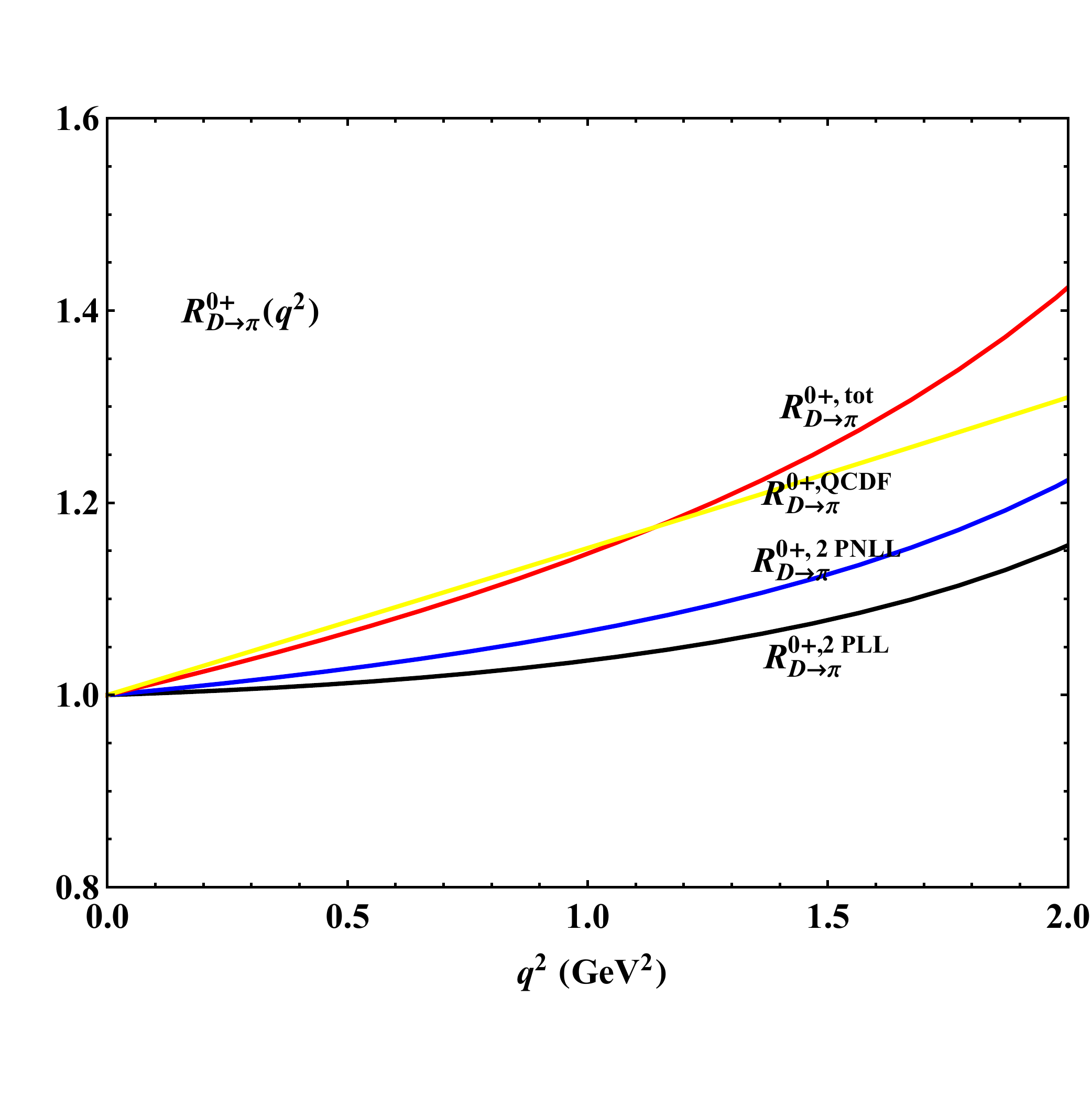}
  \includegraphics[width=0.45\textwidth]{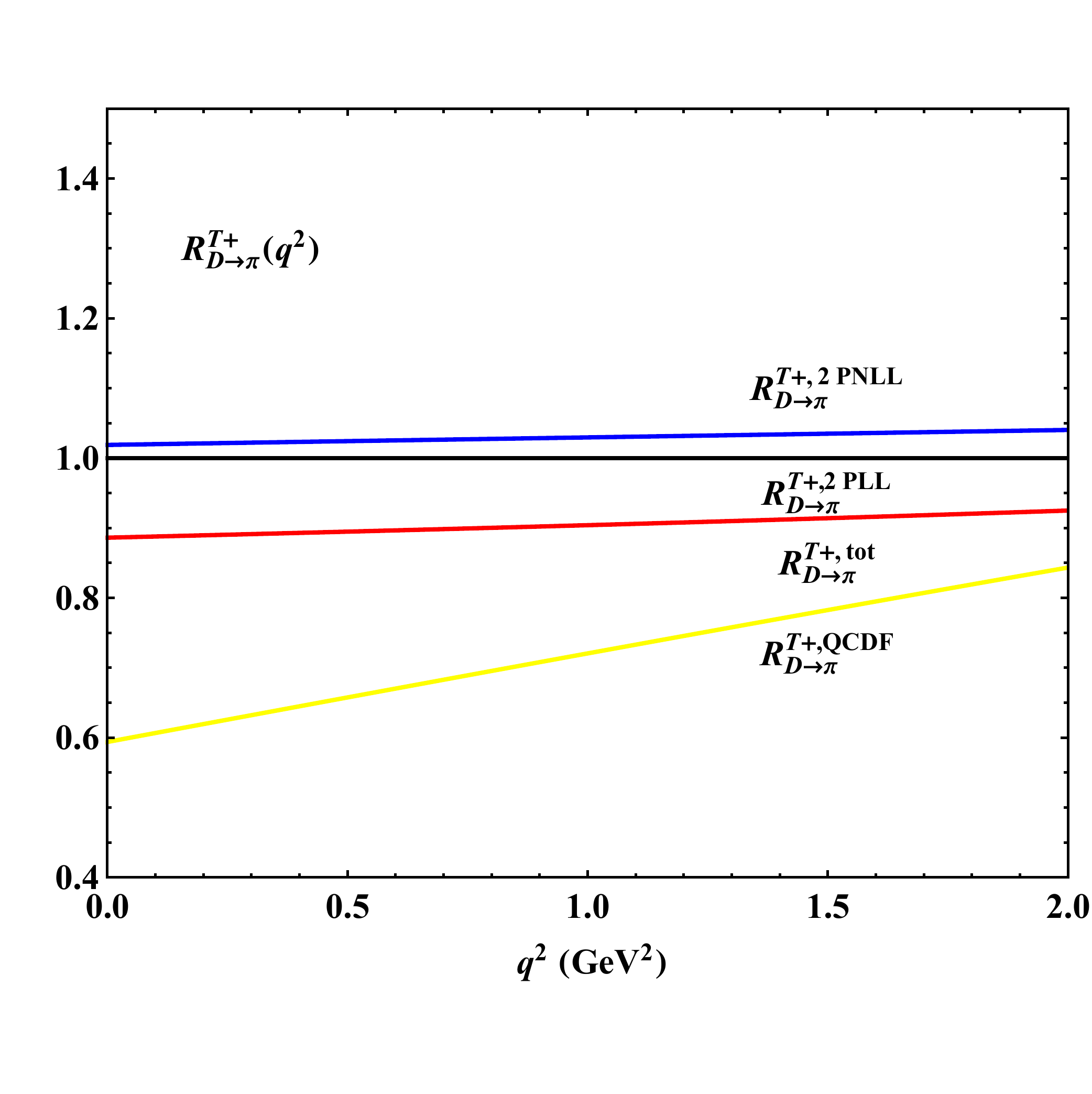}
  \caption{The large-recoil symmetry breaking effects of $D\to\pi$ form factors computed from the LCSR approach (the LL, NLL accuracy results and the tot result) and the QCD factorization  approach.}\label{figure4and5}
\end{figure}

In \cite{Charles:1998dr}, certain symmetries of the soft contributions with large final state light meson momentum have been studied, which could be broken by the perturbative QCD corrections and the higher-twist contributions. These large recoil symmetry relations of the form factors are as follows \cite{Beneke:2000wa}
\begin{eqnarray}
f_+(q^2) = \frac{m_D}{n\cdot p} \, f_0(q^2) = \frac{m_D}{m_D+m_P} \, f_T(q^2).
\label{recoil relation}
\end{eqnarray}
Within the framework of QCD factorization (QCDF), we could obtain the factorization formulae for the heavy-to-light $D$-meson form factors at one loop from \cite{Beneke:2000wa} by changing the bottom quark to a charm quark
\begin{eqnarray}
f_{D \to P}^{0}(q^2) &=& {n \cdot p \over m_D} \, f_{D \to P}^{+}(q^2)  \,
\left [  1 +  {\alpha_s \, C_F \over 2 \, \pi} \,
\left (1 -  { n \cdot p \over n \cdot p - m_D} \, \ln {n \cdot p \over m_D} \right ) \right ] \, \nonumber \\
&& +  \,  {m_D - n \cdot p \over n \cdot p } \, {\alpha_s \, C_F \over 4 \, \pi} \,
{8\, \pi^2 \, f_D \, f_P \over N_c \, m_D}  \, \int_0^1 \, d u \, { \phi_P(u, \mu) \over \bar u} \,
\int_0^{\infty} \, d \omega \, {\phi_D^{+}(\omega, \mu) \over \omega} \,,
\\
f_{D \to P}^{T}(q^2) &=& {m_D + m_P \over m_D} \, f_{D \to P}^{+}(q^2)  \,
\left [  1 +  {\alpha_s \, C_F \over 4 \, \pi} \,
\left ( \ln {m_c^2 \over \mu^2} + 2\,  { n \cdot p \over n \cdot p - m_D} \,
\ln {n \cdot p \over m_D} \right ) \right ] \, \nonumber \\
&& -  \,  {m_D + m_P \over n \cdot p } \, {\alpha_s \, C_F \over 4 \, \pi} \,
{8\, \pi^2 \, f_D \, f_P \over N_c \, m_D}  \, \int_0^1 \, d u \, { \phi_P(u, \mu) \over \bar u} \,
\int_0^{\infty} \, d \omega \, {\phi_D^{+}(\omega, \mu) \over \omega} \,,
\end{eqnarray}
where $\phi_P(u, \mu)$ denote the twist-two pseudoscalar meson LCDA.
One could obtain the form factor ratios for the semi-leptonic $D$-meson decay from (\ref{recoil relation})
\begin{eqnarray}
R_{D \to \pi}^{0 \, +}(q^2) =  {m_D \over n \cdot p} \,
{f_{D \to \pi}^{0}(q^2) \over f_{D \to \pi}^{+}(q^2)} \,,
\qquad
R_{D \to \pi}^{T \, +}(q^2) =  {m_D \over m_D + m_{\pi}} \,
{f_{D \to \pi}^{T}(q^2) \over f_{D \to \pi}^{+}(q^2)} \,,
\end{eqnarray}
and the predictions of QCDF and LCSR are shown in figure \ref{figure4and5}.
We find that predictions of the $R_{D \to \pi}^{0 \, +}(q^2)$ are consistent with each other while the prediction of $R_{D \to \pi}^{T \, +}(q^2)$ is different. Compared with the lattice prediction $R_{D \to \pi}^{T \, +}(0)=0.827$ \cite{Lubicz:2018rfs}, our result is more preferable, while the QCDF calculation is less reliable in $D$ decays.

\begin{figure}
  \centering
  \includegraphics[width=0.45\textwidth]{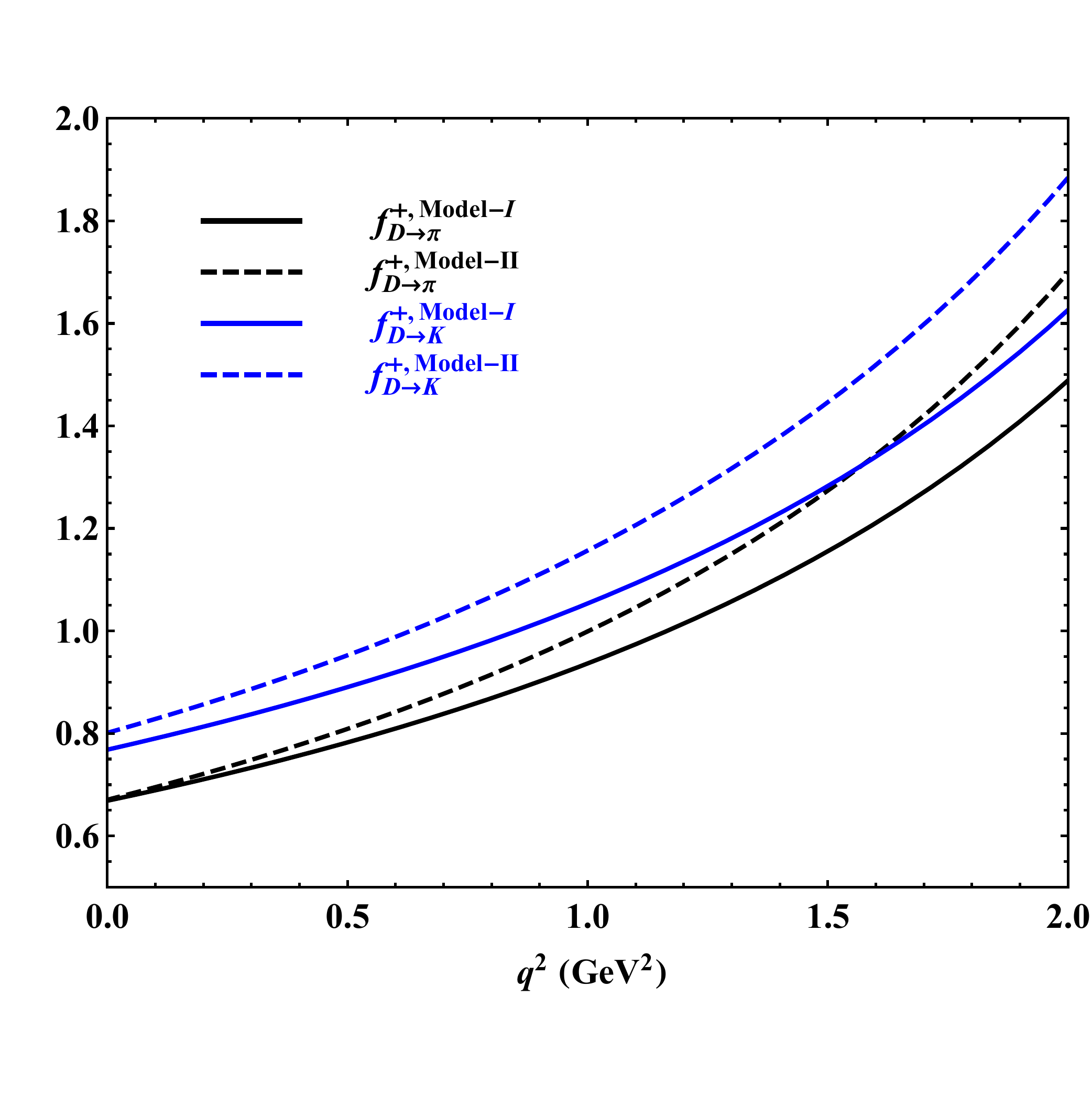}
  \includegraphics[width=0.45\textwidth]{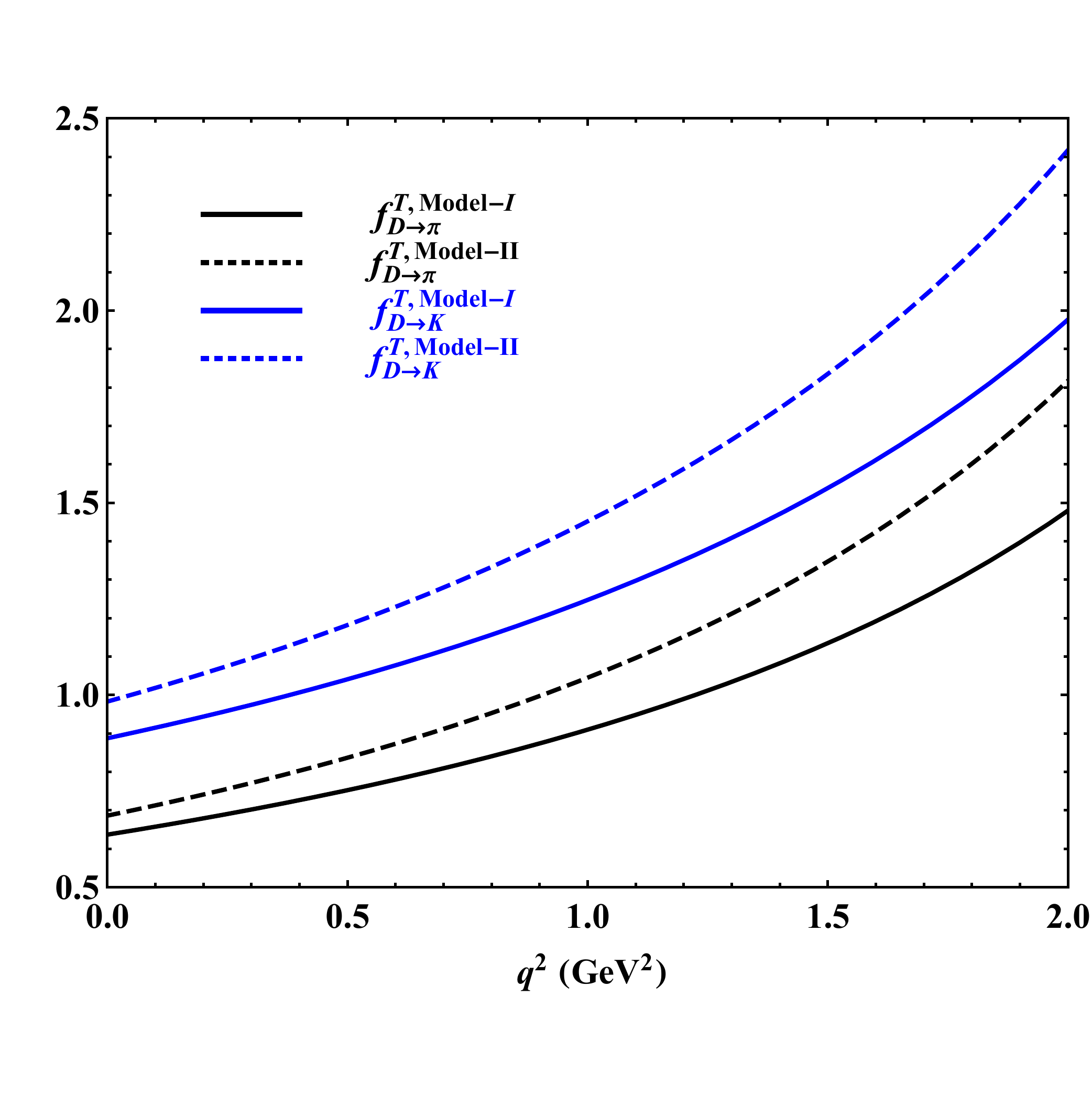}
  \caption{Dependence of $D\to\pi\,,K$ form factors on the nonperturbative models of D-meson LCDAs.}\label{figure6and7}
\end{figure}

\begin{figure}
  \centering
  \includegraphics[width=0.4\textwidth]{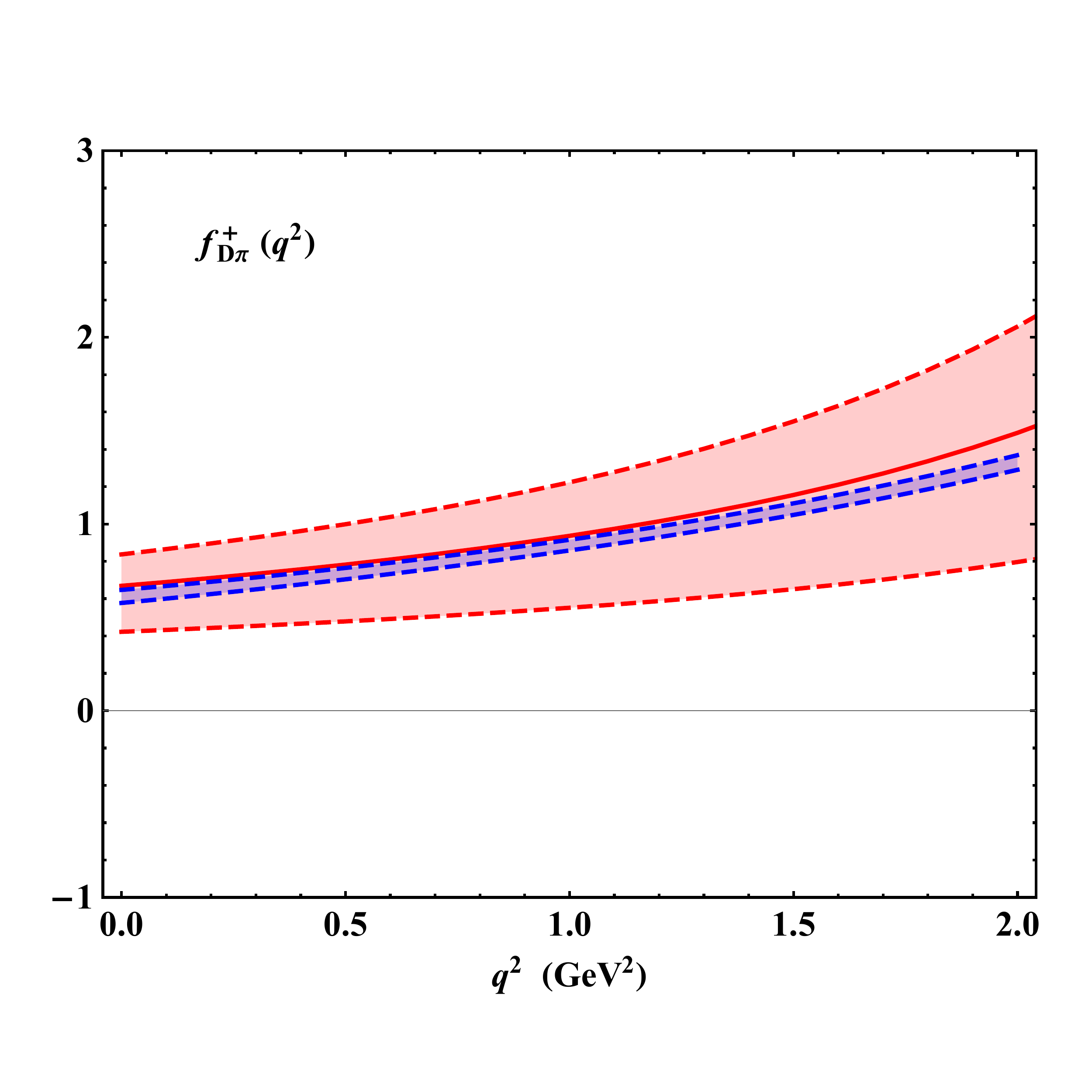}
  \includegraphics[width=0.4\textwidth]{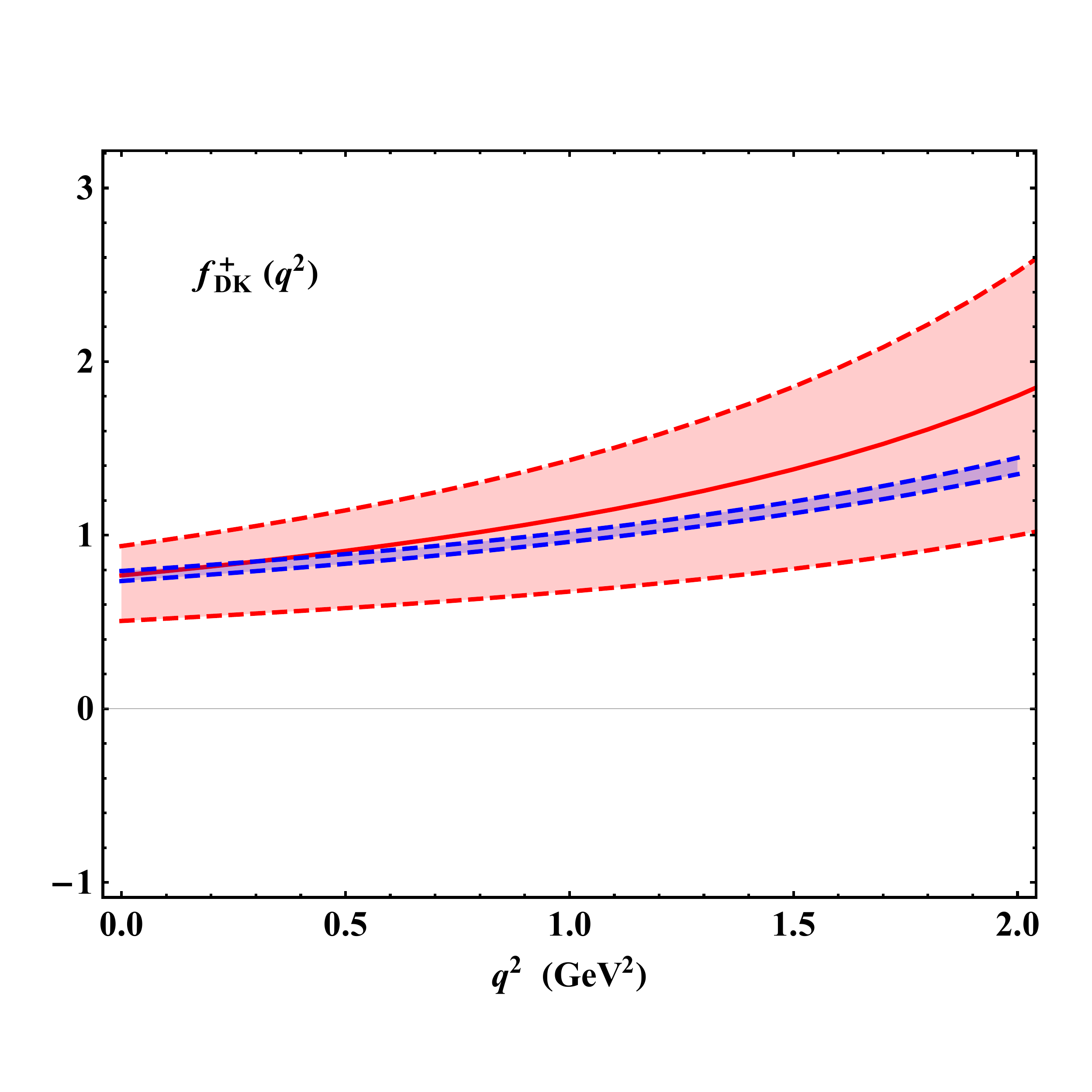}
  \includegraphics[width=0.4\textwidth]{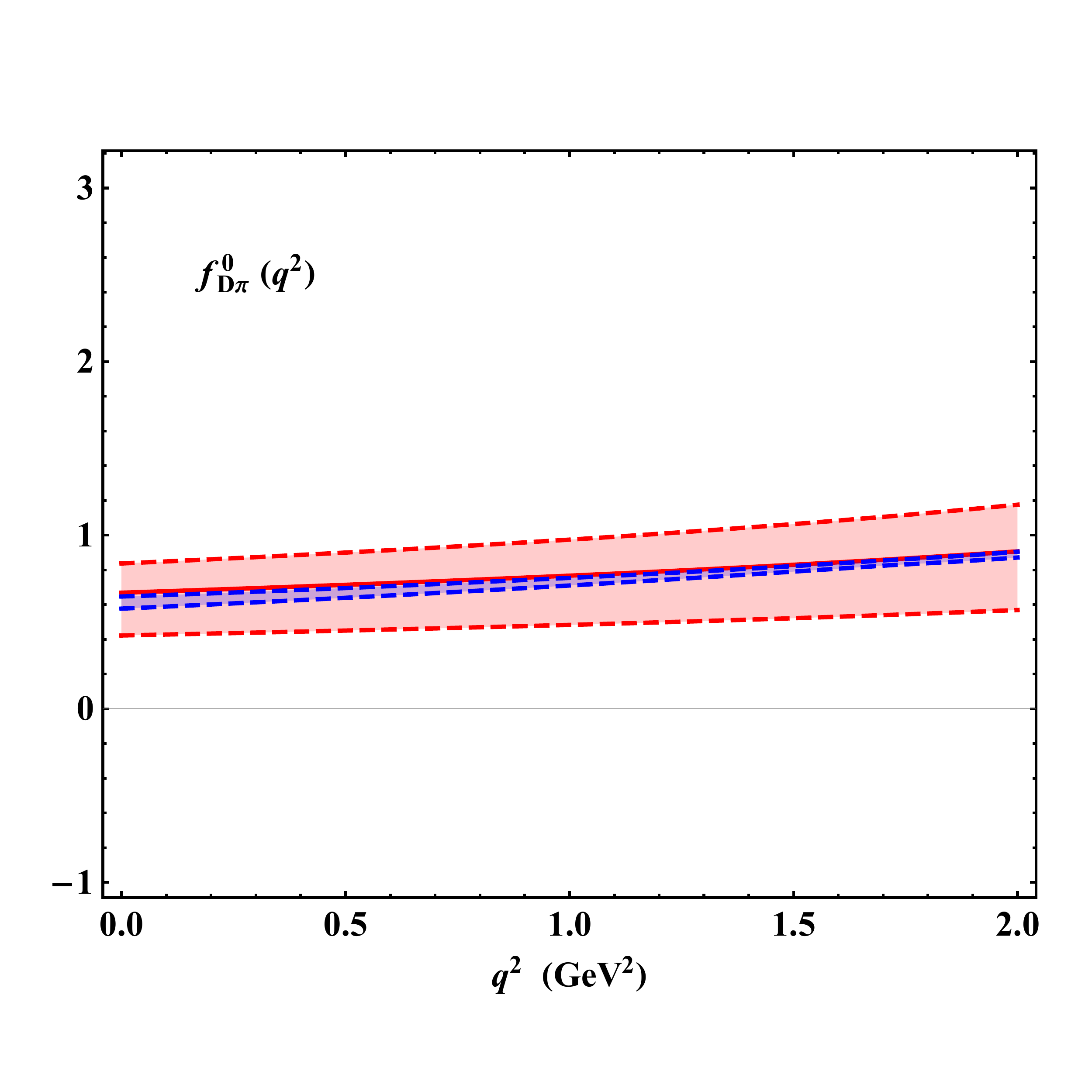}
  \includegraphics[width=0.4\textwidth]{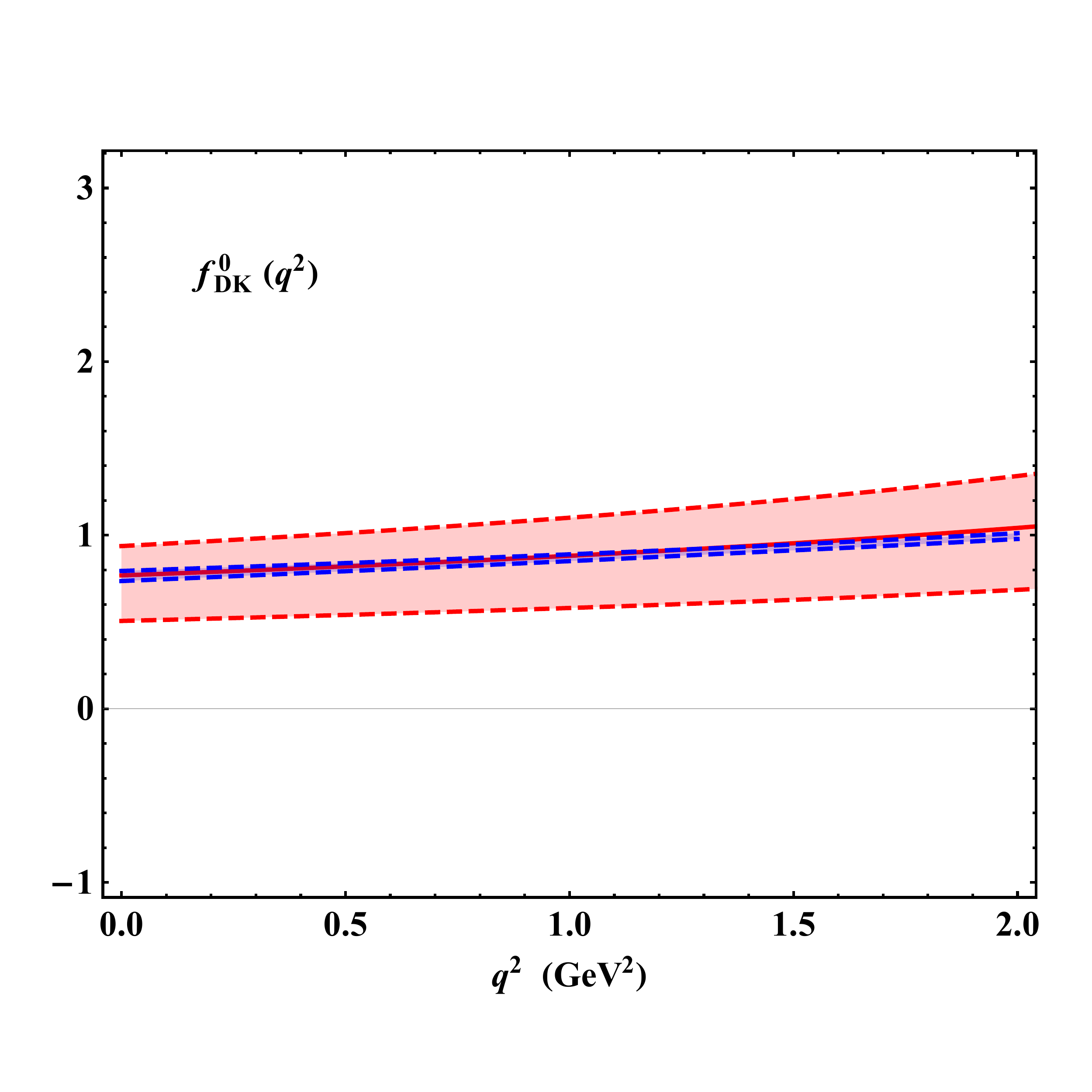}
  \includegraphics[width=0.4\textwidth]{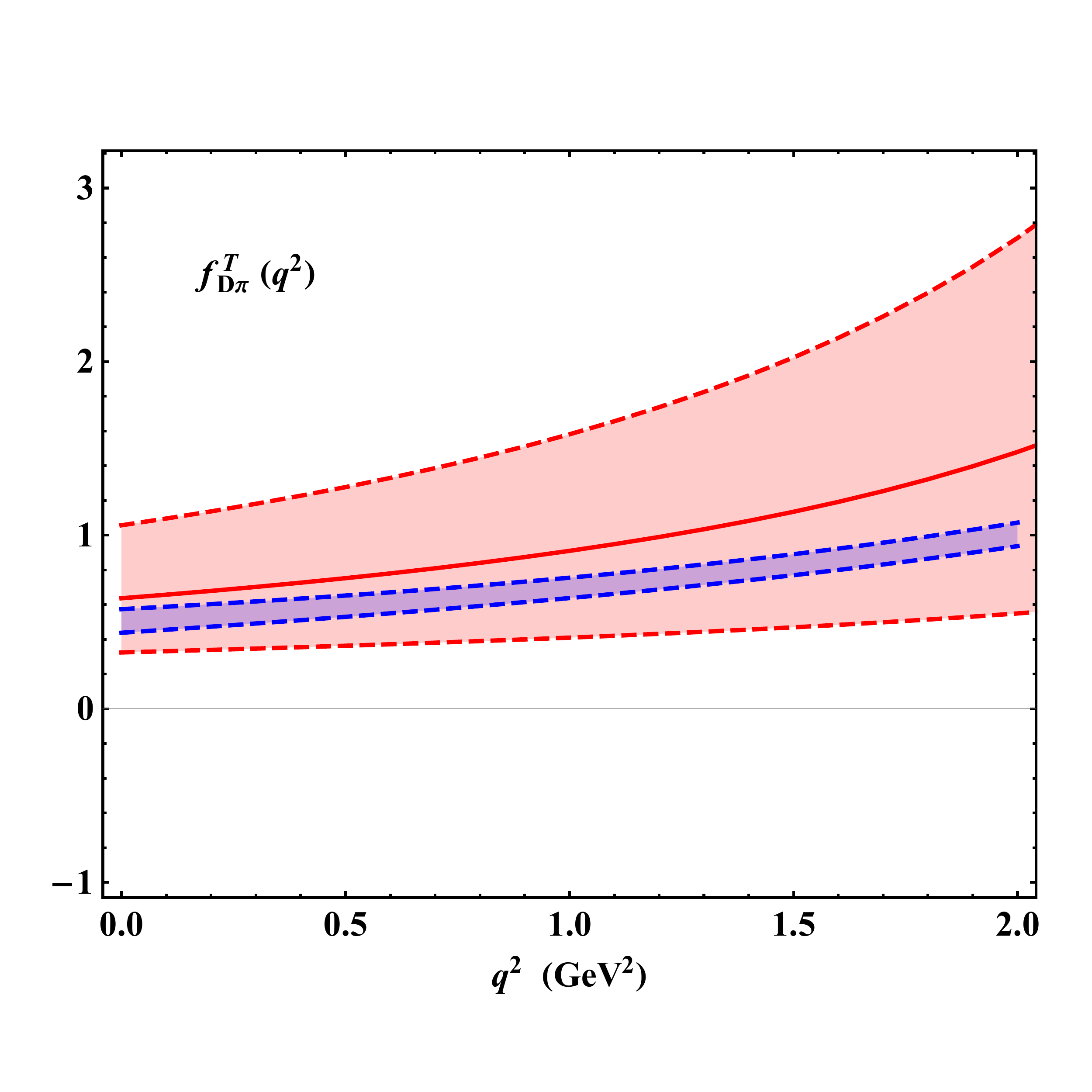}
  \includegraphics[width=0.4\textwidth]{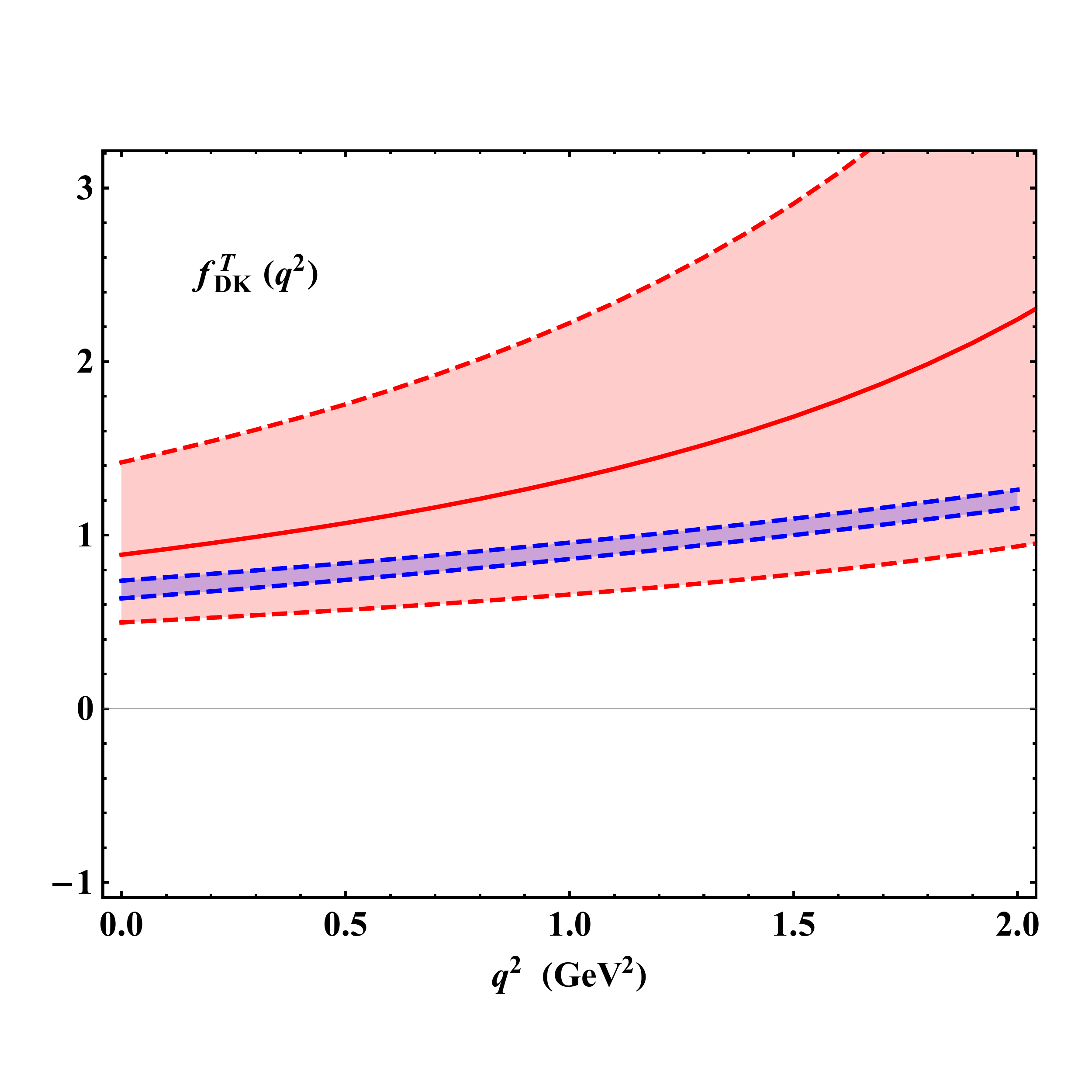}
  \caption{The momentum-transfer dependence of $D\to\pi,\,K$ form factors, where the pink bands are results from the predicted $D$-meson LCDAs and blue bands are results from ETM \cite{Lubicz:2017syv,Lubicz:2018rfs}.}\label{figure8to13}
\end{figure}

Predictions from different models are shown in figure \ref{figure6and7}, where the vector $D\to\pi$ form factor prediction at $q^2=0$ from the exponential model are fitted from \cite{Khodjamirian:2009ys}. Although the values at $q^2=0$ are almost the same, one can find that the predictions of the local duality model are more sensitive to the transfer momentum.
The dependence of form factors on the momentum transfer with uncertainties from various parameters is shown in figure \ref{figure8to13}. Comparing our predictions with the results from the Lattice QCD, we find that uncertainties of our predictions are larger and the uncertainties of the tensor form factors is more significant. Though the uncertainties of this work are significant, shapes of $D\to P$ form factors from two methods coincide with each other.
Observing dominant uncertainties from different parameters presented in table \ref{table: fitted results for the shape parameters}, we find that the  uncertainty of the NLP corrections are dominated by the inverse moment. However, the significant errors of shape parameters $b_{1,P}^i\, (i=0,T)$ are from the model dependence of $D$-meson LCDAs.
For the tensor form factors, variation of the renormalization scale $\nu$ also leads to large errors.

\begin{table}[htb]
\small
\begin{center}
\begin{tabular}{c|c|c|c|c|c|c|c|c}
\hline
\hline
Parameters                                 & Central \ value         & $\lambda_D$ & $\sigma_1$ & $M^2$  &$s_0$  & $\mu$  & $\phi_D^{\pm}(\omega)$  &$\nu$      \\  \hline
\multirow{2}{*}{$f_{D \to \pi}^{+, 0}(0)$} &  \multirow{2}{*}{0.668} & {+0.103}      & +0.026     & +0.024 &+0.018 & +0.003 & -                       & -         \\
                                           &                         & {-0.067}      & -0.028     & -0.033 &-0.020 & -0.018 & -                       & -         \\
\multirow{2}{*}{$b_{1, \pi}^{+}$}          &  \multirow{2}{*}{-0.69} & {+0.136}      & +0.049     & +0.014 &+0.041 & 0      &  0                      & -         \\
                                           &                         & {-0.107}      & -0.038     & -0.020 &-0.044 & -0.164 & -0.874                  & -         \\
\multirow{2}{*}{$b_{1, \pi}^{0}$}          &  \multirow{2}{*}{-2.10} & {+0.158}      & +0.036     & +0.031 &+0.077 & 0      & 0                       & -         \\
                                           &                         & {-0.073}      & -0.033     & -0.046 &-0.083 & -0.372 & -0.553                  & -         \\  \hline
\multirow{2}{*}{$f_{D \to \pi}^{T}(0)$}    &  \multirow{2}{*}{0.637} & {+0.126}      & +0.032     & +0.024 &+0.021 & +0.003 & -                       & +0.379    \\
                                           &                         & {-0.083}      & -0.034     & -0.034 &-0.023 & -0.031 & -                       & -0.108    \\
\multirow{2}{*}{$b_{1, \pi}^{T}$}          &  \multirow{2}{*}{-0.96} & {+0.185}      & +0.065     & +0.004 &+0.021 & 0      & 0                       & +0.227    \\
                                           &                         & {-0.141}      & -0.050     & -0.005 &-0.022 & -0.463 & -0.911                  & -0.431    \\  \hline\hline
\multirow{2}{*}{$f_{D \to K}^{+, 0}(0)$}   &  \multirow{2}{*}{0.768} & {+0.091}      & +0.024     & +0.010 &+0.007 & +0.002 & -                       & -         \\
                                           &                         & {-0.061}      & -0.025     & -0.013 &-0.008 & -0.23  & -                       & -         \\
\multirow{2}{*}{$b_{1, K}^{+}$}            &  \multirow{2}{*}{-1.025}& {+0.102}      & +0.035     & +0.044 &+0.034 & 0      & 0                       & -         \\
                                           &                         & {-0.058}      & -0.025     & -0.064 &-0.037 & -0.135 & -1.076                  & -         \\
\multirow{2}{*}{$b_{1, K}^{0}$}            &  \multirow{2}{*}{-2.103}& {+0.250}      & +0.060     & +0.078 &+0.060 & 0      & 0                       & -         \\
                                           &                         & {-0.140}      & -0.060     & -0.113 &-0.064 & -0.362 & -0.706                  & -         \\  \hline
\multirow{2}{*}{$f_{D \to K}^{T}(0)$}      &  \multirow{2}{*}{0.888} & {+0.135}      & +0.035     & +0.016 &+0.011 & +0.002 & -                       & +0.485    \\
                                           &                         & {-0.091}      & -0.037     & -0.021 &-0.012 & -0.045 & -                       & -0.138    \\
\multirow{2}{*}{$b_{1, K}^{T}$}            &  \multirow{2}{*}{-0.960}& {+0.185}      & +0.065     & +0.004 &+0.021 & 0      & 0                       & +0.227    \\
                                           &                         & {-0.141}      & -0.050     & -0.005 &-0.022 & -0.463 & -0.911                  & -0.431    \\
\hline
\hline
\end{tabular}
\end{center}
\caption{Theory predictions from model $\phi_{D,\rm I}^+(\omega,\mu_0)$ for the shape parameters and the normalizations of $D \to \pi, K$ form factors at $q^2=0$
entering the $z$ expansion with the dominant uncertainties
from variations of different input parameters.}
\label{table: fitted results for the shape parameters}
\end{table}

\begin{figure}
  \centering
  \includegraphics[width=0.45\textwidth]{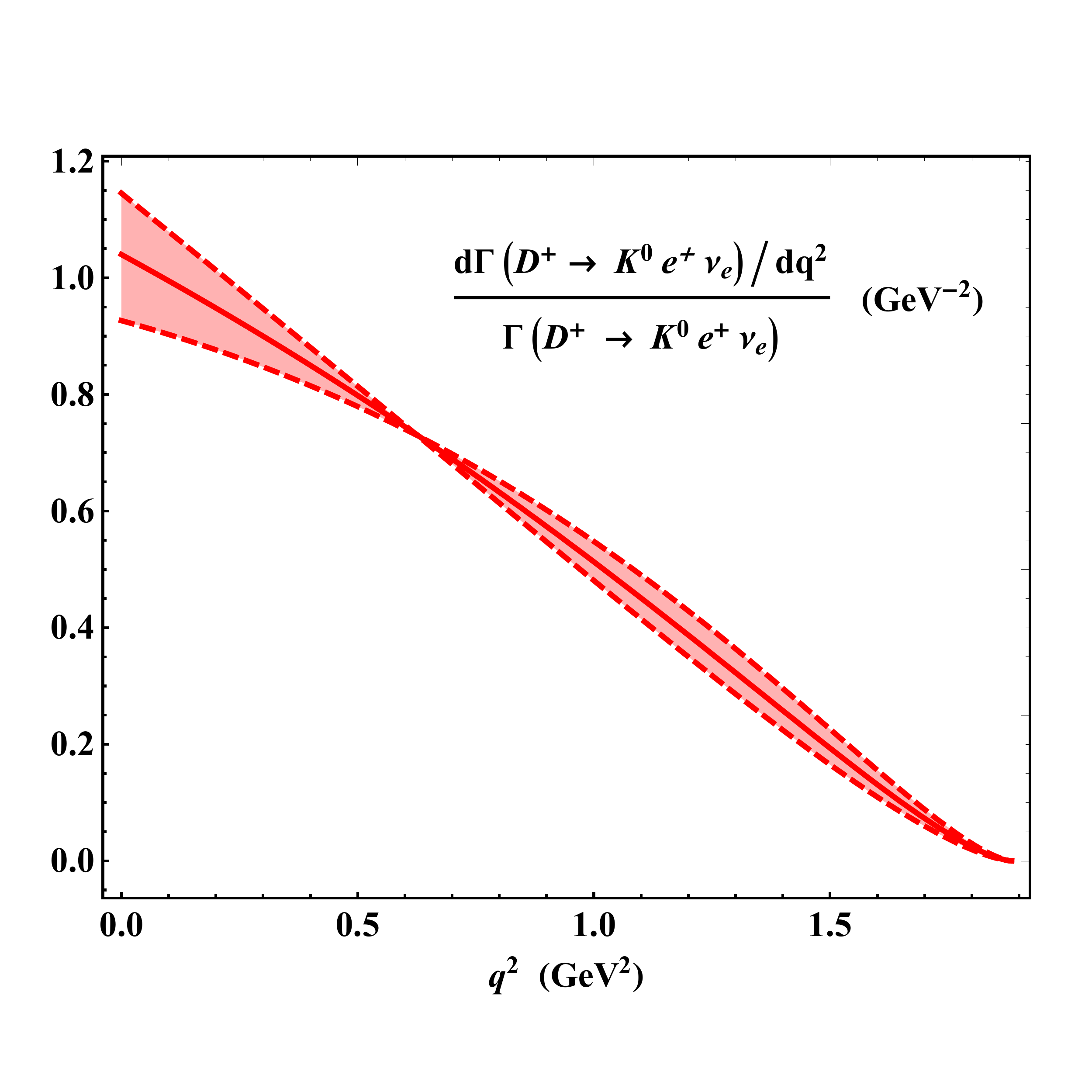}
  \includegraphics[width=0.45\textwidth]{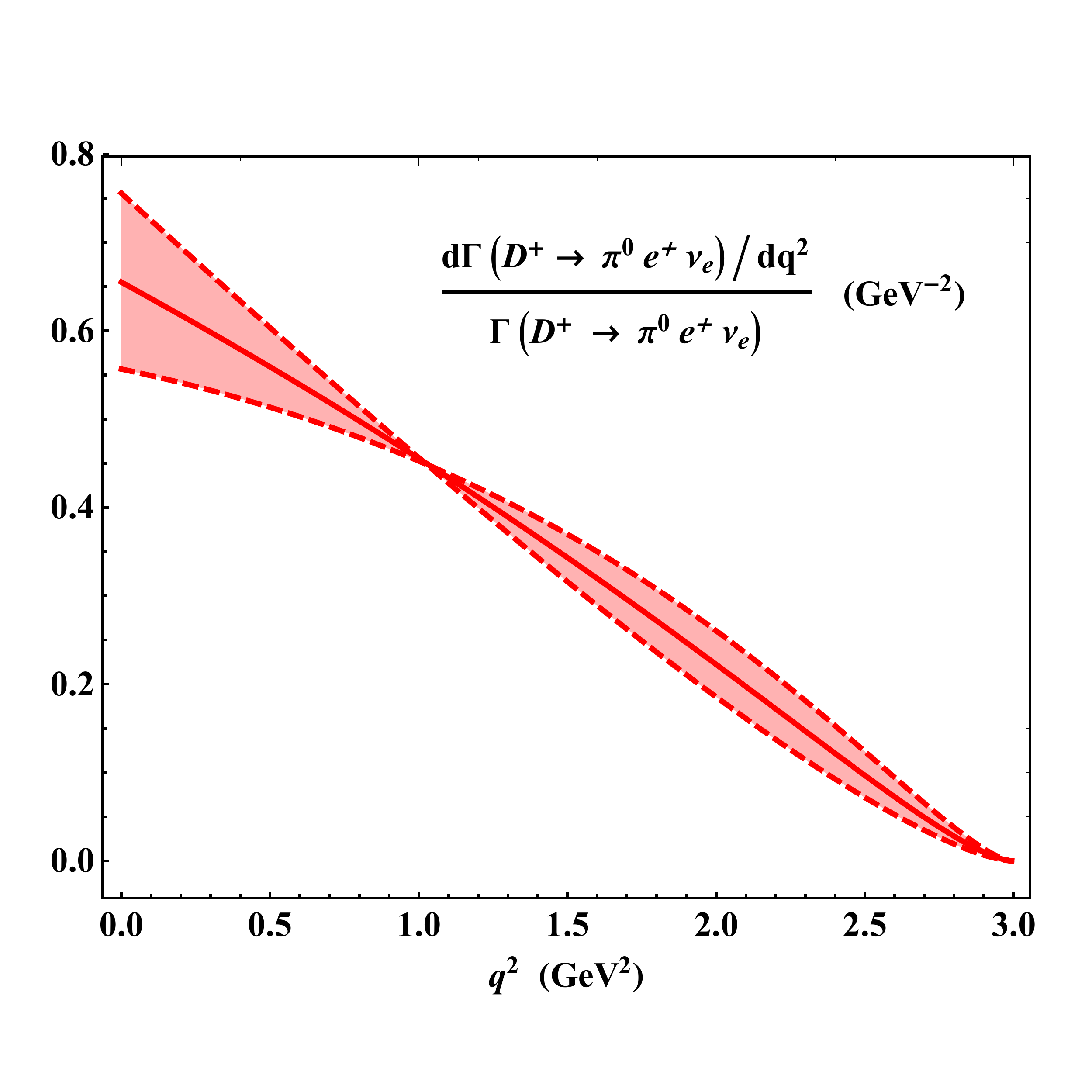}
  \caption{The normalized differential $q^2$ distributions of $D\to\pi,\,K$ decays.}\label{norm branching ratio}
\end{figure}

To extract the CKM matrix element $|V_{cq}|$, we follows the strategy presented in \cite{Khodjamirian:2006st} where the following integral is applied
\begin{eqnarray}
\Delta \zeta_{\ell}(q_1^2, q_2^2) = {1 \over |V_{cq}|^2}  \,
\int_{q_1^2}^{q_2^2} \, dq^2 \,   {d \, \Gamma (D \to \pi \ell \nu)  \over d q^2}  \, \,.
\end{eqnarray}
The differential decay rate for $D\to P\,\ell\,\nu$ is given by
\begin{eqnarray}
{d \,  \Gamma (D \to P \ell \nu)\over d q^2}
&=& {G_F^2 \, |V_{cq}|^2 \over 192 \, \pi^3 \, m_D^3} \, \lambda^{3/2}(m_D^2, m_{\pi}^2, q^2) \,
\left (  1 -{m_l^2 \over  q^2} \right )^2  \,  \left (  1 + {m_l^2 \over  2 \, q^2} \right ) \,
\bigg [ |f_{D \to \pi}^{+} (q^2)|^2  \nonumber \\
&& +  \, {3 \, m_l^2 \, (m_D^2 - m_{\pi}^2)^2 \over \lambda(m_D^2, m_{\pi}^2, q^2) \, (m_l^2 + 2\, q^2) } \,
|f_{D \to \pi}^{0} (q^2)|^2   \bigg ]  \,,
\end{eqnarray}
where $\lambda(a, b, c) = a^2 + b^2 + c^2- 2 \, ab -2 \, ac - 2 \, bc$. Employing the predictions of the $D\to\pi$ form factors in this work, we obtain the result of the $\zeta$ function after the integration of $q^2$ from 0 to 2.9 $\rm GeV^2$
\begin{eqnarray}
\Delta \zeta_{\mu}(0, 2.9 \, {\rm GeV^2}) &=&
\left ( 0.152\,{}^{+0.054}_{-0.032} \, \big|_{\lambda_D}\,{}^{+0.002}_{-0.005} \, \big|_{\sigma_1}
\,{}^{+0.0}_{-0.033} \, \big|_{M^2}
\,{}^{+0.011}_{-0.014} \, \big|_{s_0}
\,{}^{+0.023}_{-0.00} \, \big|_{\phi_D^{\pm}} \right ) \,\,\, {\rm ps}^{-1} \nonumber   \\
&=& 0.152^{+0.102}_{-0.107} \,\,\, {\rm ps}^{-1} \,,
\label{zeta integral}
\end{eqnarray}
where the second line is the result including the uncertainties from various parameters. Adopting the experimental measurement $\mathcal{B}(D^+\to\pi^0e^+\nu_e)=(3.63\pm0.08\pm0.05)\times 10^{-3}$  from the BES-III collaboration \cite{Ablikim:2017lks}, we extract the CKM matrix element $|V_{cd}|$
\begin{eqnarray}
|V_{cd}| =  0.151\,{}^{+0.091}_{-0.043} \big |_{\rm th.}\,{}^{+0.017}_{-0.02} \big |_{\rm exp.} .
\end{eqnarray}
By changing the down quark to a strange quark, and repeating the above calculation with the $q^2$ interval (0 $\sim$ 1.8) $\rm GeV^2$, one obtains
\begin{eqnarray}
\Delta \zeta_{\mu}(0, 1.8 \, {\rm GeV^2}) &=&
\left ( 0.104\,{}^{+0.027}_{-0.017} \, \big|_{\lambda_D}\,{}^{+0.001}_{-0.005} \, \big|_{\sigma_1}
\,{}^{+0.0}_{-0.023} \, \big|_{M^2}
\,{}^{+0.002}_{-0.003} \, \big|_{s_0}
\,{}^{+0.022}_{-0.00} \, \big|_{\phi_D^{\pm}} \right ) \,\,\, {\rm ps}^{-1} \nonumber   \\
&=& 0.104^{+0.06}_{-0.07} \,\,\, {\rm ps}^{-1} \,,
\label{zeta integral 2}
\end{eqnarray}
and the CKM matrix element could be extracted by adopting the experimental result of the $D\to K \, e\,\nu_e$ branching ratio $\mathcal{B}(D^+\to\bar{K}^0e^+\nu_e)=(8.6\pm0.06\pm0.15)\times 10^{-2}$ \cite{Ablikim:2017lks}
\begin{eqnarray}
|V_{cs}| =  0.89\,{}^{+0.467}_{-0.234} \big |_{\rm th.}\,{}^{+0.008}_{-0.008} \big |_{\rm exp.} .
\end{eqnarray}
Compared with the latest results from Lattice QCD and experiment (Table \ref{CKM predictions})
\begin{table}
\begin{center}
\begin{tabular}{c|c|c}
  \hline\hline
                                & $|V_{cd}|$                      & $|V_{cs}|$ \\ \hline
  ETM\cite{Lubicz:2017syv}      & 0.2221(68)                      & 1.014(25) \\ \hline
  BES-III\cite{Ablikim:2017lks} & $0.210\pm0.004\pm0.001\pm0.009$ & $0.944\pm0.005\pm0.015\pm0.024$ \\ \hline
  BES-III\cite{Ablikim:2018evp} & -                               & $0.955\pm0.005\pm0.004\pm0.024$ \\\hline
\end{tabular}
\end{center}
\caption{The CKM matrix element values from ETM and BES-III collaborations.}
\label{CKM predictions}
\end{table}
our prediction of $|V_{cd}|$ is much lower than others, which indicates that the predicted form factors is too large. This result may due to that fact that many important power suppressed contributions such as the power suppressed term in the heavy-to-light current are not taken into account. A more careful study on the  power suppressed contribution is needed.
In Fig. \ref{norm branching ratio}, we present the normalized differential $q^2$ distributions of $D\to\pi,\,K$. Considering no experimental data is provided, we hope relevant experiments could be conducted in the future.

\section{Conclusion}

The $D\to P$ transition form factors are the fundamental nonperturbative parameters in the semi-leptonic and non-leptonic $D$ decays, thus it is of great importance in the determination of CKM matrix elements $V_{cq}$. Employing the method of LCSR, we calculated the $D \to \pi, \,K $ form factors, including the NLL resummation of the leading order contribution and  the higher-twist contributions from the two-particle and three-particle $D$-meson LCDAs at tree level. The numerical results indicate that the corrections from the two-particle higher-twist contributions are of $\mathcal{O}(27\%\sim36\%)$ while corrections from three-particle higher-twist contributions are tiny. The SU(3) flavor symmetry breaking and the large recoil symmetry breaking effects are also studied in this work. Especially, the predicted   SU(3) flavor symmetry breaking effect  of vector (scalar) and tensor are $R_{\rm SU(3)}^{+}$ ($R_{\rm SU(3)}^{0}$)=1.12 and $R_{\rm SU(3)}^T$=1.39, and our predicted $R_{\rm SU(3)}^T$ result is consistent with the Lattice result 1.36 \cite{Lubicz:2018rfs}. The large recoil symmetry which holds for soft form factors  can be broken by three-particle higher-twist corrections. When comparing the predictions of the QCDF and LCSR for the large recoil symmetry breaking effect, we found the QCDF prediction of $R_{D\to\pi}^{T+}$ is less preferable and our result is more close to the lattice data.

As some parameters such as the inverse moment and the shape of the $D$-meson LCDA are not well determined, the uncertainties of our predictions of form factors are large. To study the uncertainties from different variables, we adjusted the inverse moment to reproduce the vector $D\to\pi$ form factor result from the pion LCSR.
The predicted form factors were further applied to extract the CKM matrix elements.
By utilizing the measured branching ratios and doing the integration of $\zeta$ function, we obtain the predicted $|V_{cd}|=0.151\,{}^{+0.091}_{-0.043} \big |_{\rm th.}\,{}^{+0.017}_{-0.02} \big |_{\rm exp.}$ and $|V_{cs}|= 0.89\,{}^{+0.467}_{-0.234} \big |_{\rm th.}\,{}^{+0.008}_{-0.008} \big |_{\rm exp.}$. When comparing our results with the experimental and the lattice results, we found the prediction of $|V_{cs}|$ is close to them. Though $|V_{cd}|$ of this work is lower than results from the BES-III and the ETM collaborations, this result is still reasonable within the uncertainties of the predicted CKM matrix elements, and a more careful study with profound investigations of the power corrections is needed to reduce the uncertainties.
We can further investigate the subleading power contributions not considered in this paper, such as the subleading power correction from heavy quark expansion in HQET, the subleading power corrections from the quark propagator expansion at tree level, and the four-particle D-meson LCDA corrections.

\subsection*{Acknowledgements}
This work is supported in part by the National Natural Science Foundation of China with Grant No. 11675082 and 11735010, and the Natural Science Foundation of Tianjin with Grant No. 19JCJQJC61100. The author would like to thank Yu-Ming Wang for illuminating discussions.

\appendix

\section{Evolution functions}
We collect evolution functions $U_1(E_\gamma,\mu_h,\mu)$ and $U_2(E_\gamma,\mu_h,\mu)$ from \cite{Beneke:2011nf}, and $U_2(E_\gamma,\mu_h,\mu)$ is obtained by setting the cusp anomalous dimension to zero, details could be found in this reference.
\begin{eqnarray}
U_1(E_\gamma,\mu_h,\mu) &=&
\exp\left(\,\int_{\alpha_s(\mu_h)}^{\alpha_s(\mu)} d\alpha_s\,
\left[
\frac{\gamma(\alpha_s)}{\beta(\alpha_s)} +
\frac{\Gamma_{\rm cusp}(\alpha_s)}{\beta(\alpha_s)}
\left(
\ln\frac{2 E_\gamma}{\mu_h} -
\int_{\alpha_s(\mu_h)}^{\alpha_s}
\frac{d\alpha_s^\prime}{\beta(\alpha_s^\prime)}
\right)
\right]\right)
\nn \\[0.4cm]
&& \hspace*{-2.2cm}
=\,\exp\left(\,
-\frac{\Gamma_0}{4\beta_0^2} \left(
\frac{4\pi}{\alpha_s(\mu_h)}\left[\ln r-1+\frac{1}{r}\right]
-\frac{\beta_1}{2 \beta_0} \,\ln^2 r
+\left(\frac{\Gamma_1}{\Gamma_0}-\frac{\beta_1}{\beta_0}\right)
\left[r-1-\ln r\right]\right)
\right)
\nn \\[0.2cm]
&& \hspace*{-1.7cm}
\times\,\left(\frac{2 E_\gamma}{\mu_h}\right)^
{-\frac{\Gamma_0}{2\beta_0} \ln r}
 r^{-\frac{\gamma_0}{2\beta_0}}
\times \Bigg[1 - \frac{\alpha_s(\mu_h)}{4\pi}\,\frac{\Gamma_0}{4\beta_0^2}
\,\bigg(\frac{\Gamma_2}{2\Gamma_0} \left[1-r\right]^2
+\frac{\beta_2}{2\beta_0} \left[1-r^2+2 \ln r\right]
\nn \\[0.2cm]
&& \hspace*{-0.7cm}
-\,\frac{\Gamma_1\beta_1}{2\Gamma_0\beta_0}
\left[3-4 r+r^2+2 r \ln r\right]
+\frac{\beta_1^2}{2\beta_0^2} \left[1-r\right]\left[1-r-2\ln r\right]
\bigg)
\nn \\[0.2cm]
&& \hspace*{-0.7cm}
+\,\frac{\alpha_s(\mu_h)}{4\pi}\left(
\ln\frac{2E_\gamma}{\mu_h}
\left(\frac{\Gamma_1}{2\beta_0}-\frac{\Gamma_0\beta_1}{2\beta_0^2}\right)
+\frac{\gamma_1}{2\beta_0}-\frac{\gamma_0\beta_1}{2\beta_0^2}\right)
\left[1-r\right] + {\cal O}(\alpha_s^2)\Bigg]
\label{U1}
\end{eqnarray}

The QCD evolution factor $U_3(\nu_{h}, \nu)$  is given by \cite{Lu:2018cfc}
\begin{eqnarray}
U_3(\nu_{h}, \nu) &=& {\rm Exp}  \bigg [ \int_{\alpha_s(\nu_{h})}^{\alpha_s(\nu)} \,
d \alpha_s \, \frac{\gamma_T(\alpha_s)}{\beta(\alpha_s)} \bigg ] \, \nonumber \\
&=& z^{- \frac{\gamma_T^{(0)}}{2 \,\beta_0}} \bigg [1+ \frac{\alpha_s(\nu_{h})}{4 \pi}  \,
\left (  {\gamma_T^{(1)} \over 2 \, \beta_0} - {\gamma_T^{(0)} \, \beta_1 \over 2 \, \beta_0^2 } \right ) (1-z)
+{\cal O}(\alpha_s^2) \bigg ]\,,
\end{eqnarray}
with $z=\alpha_s(\nu)/\alpha_s(\nu_h)$. The anomalous dimension $\gamma_T(\alpha_s)$ for the tensor current
at the two-loop accuracy is \cite{Bell:2010mg}
\begin{eqnarray}
\gamma_T(\alpha_s) &=& \sum_{n=0}^{\infty} \, \left ( {\alpha_s(\mu) \over 4 \, \pi} \right )^{n+1} \,
\gamma_T^{(n)} \,, \qquad \gamma_T^{(0)} = - 2\, C_F  \,, \nonumber \\
\gamma_T^{(1)} &=&   C_F \, \left [ 19 \, C_F \,  -{257 \over 9} \, C_A
+  {52 \over 9} \, n_f \, T_F \right ]\,.
\end{eqnarray}

\section{Expressions of $\rho_{i, \rm{LP}}^{(3P)}$ and $\rho_{i, \rm{NLP}}^{(3P)}$}
The expressions of $\rho_{i, \rm{LP}}^{(3P)}$ and $\rho_{i, \rm{NLP}}^{(3P)}$
($i=n\,, \bar n \,, T$) are evaluated in \cite{Lu:2018cfc}, and they are given by
\begin{eqnarray}
\rho_{\bar n, \rm{LP}}^{(3P)} &=& (1-2 \, u) \, \left [ X_A - \Psi_A- 2 \, Y_A\right ]
-\tilde{X}_A  - \Psi_V  + 2\, \tilde{Y}_A  \,, \nonumber \\
\rho_{\bar n, \rm{NLP}}^{(3P)} &=& 2 \, \left [ \Psi_A - \Psi_V \right ]
+ 4 \, \left [ W + Y_A +  \tilde{Y}_A  - 2 \, Z \right ]  \,, \nonumber \\
\rho_{n, \rm{LP}}^{(3P)} &=& 2\, (u-1) \, \left ( \Psi_A + \Psi_V \right ) \,, \nonumber \\
\rho_{n, \rm{NLP}}^{(3P)} &=& \left ( \Psi_A - \Psi_V \right )
- \, \left [ X_A +  \tilde{X}_A  - 2 \, Y_A - 2\, \tilde{Y}_A  \right ]  \,, \nonumber \\
\rho_{T, \rm{LP}}^{(3P)} &=&  (1-2 \,u) \, \left ( \Psi_V + X_A  - 2 \, Y_A \right )
+ \Psi_A -  \tilde{X}_A + 2\, \tilde{Y}_A \,, \nonumber \\
\rho_{T, \rm{NLP}}^{(3P)} &=& \left ( \Psi_A - \Psi_V + X_A +  \tilde{X}_A \right )
+ 2 \, \left [ 2\, W +  Y_A  + \, \tilde{Y}_A  - 4 \, Z \right ]  \,,
\end{eqnarray}

\section{Fourier transformation of the nontrivial relations of $D$-meson LCDAs}
One could obtain the $D$-meson LCDAs by changing the bottom quark to a charm quark due to the heavy quark symmetry.
Following \cite{Kawamura:2001jm,Braun:2017liq}, we could express the twist 4 DA $g_D^+$ and twist 5 DA $g_D^-$ in terms of three particle DAs. Applying the operator identities (\ref{identity}), one obtains
\begin{subequations}
\label{KKQT}
\begin{align}
\hspace*{-0.5cm}  \Big[z\frac{d}{dz}+1\Big]\Phi_-(z) &=  \Phi_+(z)  + 2 z^2  \int_0^1\! udu\,\Phi_3(z,uz)\,,
\label{KKQT1}
\\
2 z^2  \mathrm{G}_+(z) & =
-  \Big[ z \frac{d}{dz} - \frac12  + i z \bar \Lambda \Big] \Phi_+(z)
-  \frac{1}{2}\Phi_-(z)
- z^2  \int_0^1\! \bar udu\,{\Psi}_4(z,uz)\,,
\label{KKQT2}
\\
 2 z^2 \mathrm{G}_-(z)
&= -  \Big[ z \frac{d}{dz} - \frac12  + i z \bar \Lambda \Big] \Phi_-(z) - \frac12  \Phi_+(z)
- z^2  \int_0^1\! \bar udu\,{\Psi}_5(z,uz)\,,
\label{KKQT3}
\\
 \Phi_-(z)
&= \left(z \frac{d}{dz}+1 + 2i z \bar \Lambda  \right) \Phi_+(z) +
2 z^2 \int_0^1\! du\,  \Big[ u \Phi_4(z,uz) + {\Psi}_4(z,uz)\Big],
\label{KKQT4}
\end{align}
\end{subequations}
where
\begin{align}
  \mathrm{G}_\pm(z,\mu) &= \int\limits_0^\infty d\omega \, e^{-i\omega z}g_\pm(\omega,\mu)
\end{align}
and
\begin{align}
   \bar\Lambda = m_D -m_c\,.
\end{align}
Implementing the definition of the momentum space distributions
\begin{align}
  \Psi_A(z_1,z_2)=\int_{0}^{\infty}d\omega_1\int_{0}^{infty}d\omega_2e^{-i\omega_1z_1-i\omega_2z_2}\psi_A(\omega_1,\omega_2)\,,
\end{align}
and doing Fourier transformation, one could obtain (\ref{the first EOM}), (\ref{the second EOM}), (\ref{the third EOM}) and (\ref{the fourth EOM}).

\end{document}